\documentclass[aps,pra,twocolumn,groupedaddress,showpacs]{revtex4-1}

\usepackage{bm}
\usepackage{amsmath}
\usepackage[usenames,dvipsnames]{xcolor}
\usepackage{graphicx}
\usepackage[normalem]{ulem}
\usepackage{enumitem}
\usepackage{booktabs}

\newcommand{\vcn}[1]{\hat{#1}}
\newcommand{\vcc}[1]{{\boldsymbol{#1}}}
\newcommand{\vc}[1]{\underline{#1}}

\newcommand{\mat}[1]{\underline{\underline{#1}}}
\providecommand{\bra}[1]{\ensuremath{\left\langle \,#1\, \right\rvert}}

\providecommand{\ket}[1]{\ensuremath{\left\lvert \,#1\, \right\rangle}}

\newcommand{\abs}[1]{\lvert #1 \rvert}
\newcommand{\expo}[1]{\mathrm{e}^{#1}}
\renewcommand{\eqref}[1]{(\ref{#1})}
\renewcommand{\Im}{\ensuremath{\mathrm{Im}}}
\renewcommand{\Re}{\ensuremath{\mathrm{Re}}}
\newcommand{\atan}{\ensuremath{\mathrm{atan}}}
\newcommand{\imag}{\ensuremath{\mathrm{i}}} 

\newcommand{\pauli}{\ensuremath{\boldsymbol{\sigma}^\mathrm{P}}}
\newcommand{\paulix}{\ensuremath{\sigma_x^\mathrm{P}}}
\newcommand{\pauliy}{\ensuremath{\sigma_y^\mathrm{P}}}
\newcommand{\pauliz}{\ensuremath{\sigma_z^\mathrm{P}}}
\newcommand{\td}{\ensuremath{\mathrm{d}}}

\newcommand\SmallMatrix[1]{{%
		\tiny\arraycolsep=0.3\arraycolsep\ensuremath{\begin{pmatrix}#1\end{pmatrix}}}}

\renewcommand{\eqref}[1]{(\ref{#1})}
\newcommand{\teqref}[1]{Eq.~\eqref{#1}}
\newcommand{\eg}{e.g.\@ }
\newcommand{\ie}{i.e.\@ }
\newcommand{\cf}{cf.\@ }
\newcommand{\etal}{\textit{et al.\@ }}

\newcommand{\sqa}{\ensuremath{\vcn{s}}}
\newcommand{\asqks}{\ensuremath{a^2_{\vcc{k} \sqa}}}
\newcommand{\bsqks}{\ensuremath{b^2_{\vcc{k} \sqa}}}
\newcommand{\bsq}{\ensuremath{b^2_{\sqa}}}

\begin{document}


\title{Fermi surfaces, spin-mixing parameter, and colossal anisotropy of spin relaxation in transition metals from \textit{ab initio} theory}


\author{Bernd Zimmermann}
\email{be.zimmermann@fz-juelich.de}
\author{Phivos Mavropoulos}
\author{Nguyen H. Long}
\author{Christian-Roman Gerhorst}
\author{Stefan Bl\"{u}gel}
\author{Yuriy Mokrousov}
\affiliation{Peter Gr\"{u}nberg Institut and Institute for Advanced Simulation, Forschungszentrum J\"{u}lich and JARA, 52425 J\"{u}lich, Germany}


\date{\today}

\begin{abstract}
  The Fermi-surfaces and Elliott-Yafet spin-mixing parameter (EYP) of several elemental metals are studied by \emph{ab initio} calculations. We focus first on the anisotropy of the EYP as a function of the direction of the spin-quantization axis [Phys.~Rev.~Lett.\ \textbf{109}, 236603 (2012)]. We analyze in detail the origin of the gigantic anisotropy in $5d$ hcp metals as compared to $5d$ cubic metals by band-structure calculations and discuss the stability of our results against an applied magnetic field. We further present calculations of light (4$d$ and 3$d$) hcp crystals, where we find a huge increase of the EYP anisotropy, reaching colossal values as large as $6000\%$ in hcp Ti. We attribute these findings to the reduced strength of spin-orbit coupling, which promotes the anisotropic spin-flip hot loops at the Fermi surface.
  In order to conduct these investigations, we developed an adapted tetrahedron-based method for the precise calculation of Fermi surfaces of complicated shape and accurate Fermi-surface integrals within the full-potential relativistic Korringa-Kohn-Rostoker Green-function method.
\end{abstract}

\pacs{72.25.Rb,72.25.Ba,76.30.Pk,75.76.+j}



\maketitle


\section{Introduction}

The Fermi surface (FS) is of special importance for many properties of metals \cite{Sprongford_FermiSurface}. The low-energy transitions between occupied and unoccupied states close to the Fermi energy govern electronic \cite{ziman1972principles,Mertig_Transport} and spin-transport properties \cite{Spintronics_Zutic}, as well as response functions and their instabilities. Moreover, the Fermi surface takes a special role for quasiparticle excitations in Landau Fermi-liquid theory, as their lifetime tends towards infinity as the energy approaches the Fermi level \cite{philipps_AdvSolStatePhys}.
Purely the area of the Fermi surface already influences the density of states, and thus determines the low-temperature specific heat, as well as the ferromagnetic instability through the Stoner criterion. Some more important physical effects are determined merely by the shape of the FS. For example, the extremal orbits of the Fermi surface determine the de Haas-van Alphen oscillations \cite{Kittel_solidstatephysics}. The Fermi wave vector directly influences the period of Friedel oscillations and the Ruderman-Kittel-Kasuya-Yosida-type exchange interaction. Some more advanced properties of the Fermi surface originate from nesting and have ramifications in charge- or spin-density waves \cite{Overhauser.CDW,RevModPhys.SDW.Cr}, the shape memory effect \cite{PAPACONSTANTOPOULOS198293}, a focusing effect of Friedel oscillations around impurities \cite{Weismann.focussingEffect,Mohammed:QWS}, or superconductivity \cite{0953-2048-14-4-201}.

The shape of a Fermi surface can vary from a simple sphere for a homogeneous electron gas up to very complex shapes, which is especially the case for transition-metal elements due to the presence of dense $d$-electron bands at the Fermi level \cite{note:fermisur-database}. They often exhibit many intertwined sheets and possible crossings or anti-crossings (corresponding respectively to degenerate electron states or lifted degeneracies in the band structure). The lifting of these degeneracies is frequently caused by spin-orbit coupling (SOC) \cite{Pientka.2012}, and the precise determination of resulting small anti-crossings of Fermi-surface sheets is often crucial for the correct description of spin-orbit effects in metals.

Spin-orbit coupling manifests itself in various effects of high fundamental and technological relevance, including anisotropy effects and spin-dependent transport phenomena. The former class includes the magneto-crystalline anisotropy energy (MAE) and anisotropic magnetoresistance (AMR). Examples for the latter are the anomalous as well as the direct and inverse spin-Hall effects (AHE, SHE and ISHE) \cite{AHE_Nagaosa,skewscattering_zimmermann,Dyakonov-Perel:SHE,Hirsch:spin-hall,Kato:spin-hall:exp}, which lie at the heart of modern spintronics for spin-current creation and detection. Moreover, the important phenomenon of spin-relaxation determines the time scale on which an excited spin population, which is for example created by an injected spin-polarized current, equilibrates, and is therefore a crucial parameter for the design of spintronic devices.

Evidently, a sufficiently long spin-relaxation time $T_1$ is required if information encoded in the orientation of the electron spin shall be transported across a device by means of a spin-polarized current, because this current has basically decayed after this time \cite{Spintronics_Zutic}. On the other hand, a short spin-relaxation time may be required in ultrafast demagnetization dynamics \cite{Carva_Oppeneer.2011}, where an excited spin-population (\eg by means of a laser) is used to transfer energy quickly into other degrees of freedom, \eg into the lattice by electron-phonon coupling. Usually, for the two limiting cases of long or short $T_1$, two different materials are needed. However, recently a novel anisotropy of spin-relaxation as a function of the spin-direction of the spin-population was discovered \cite{Zimmermann:Spinrelaxation:2012}. This anisotropy can be gigantic (as large as 830\% in Hf), and allows for an adjustment of the spin-relaxation time within the same material, just by changing the polarization direction of the excited spin population.

From a numerical point of view, the accurate determination of complex Fermi surfaces and the precise calculation of Fermi-surface integrals represent a true challenge. A widely used concept of dividing the irreducible Brillouin zone into tetrahedra and interpolating the integrand by a linear function (linear tetrahedron method) was first proposed by Lehmann and Taut \cite{Lehmann:tetramerhod}, and refinements have led to higher computational efficiency and accuracy \cite{Bloechl:tetramethod,Eiguren_Fermisurface_Helmholtz}. Like most integration methods for the reciprocal space, these formulations rely on the knowledge of band energies $\epsilon_{i} (\vcc{k})$ at the vertices of the tetrahedra (with the Bloch vector $\vcc{k}$ and band index $i$; see \cite{Eyert_ASW} for an overview). However, in the Korringa-Kohn-Rostoker Green function (KKR-GF) method, which has many advantages over basis-set based methods (\eg for the inclusion of disorder, scattering properties and corresponding transition rates \cite{Mertig_Transport,Zeller_impurities,dederichs:KKR-NIC,Ebert:review:KKR,PhysRevLett.104.186403}), the band structure is given by an implicit relation between $\epsilon_{i}$ and $\vcc{k}$. An adapted method is needed for the KKR formalism, which relies on the search for roots of the KKR-matrix eigenvalues, $\lambda_i(\vcc{k},\epsilon)=0$. It was initially formulated for the atomic sphere approximation (ASA) and on a tetrahedral mesh by Zahn \cite{Zahn.phd}. However, complications arise from the inclusion of non-spherical parts into the potential. The full-potential treatment becomes especially important for surfaces and layered systems, as well as in magnetic bulk crystals with spin-orbit coupling.

In the present paper, we begin by presenting a robust method for calculating the Fermi surface based on an adapted tetrahedron method within the relativistic full-potential KKR-GF formalism, which enables the determination of Fermi-surfaces of most complicated shape. We apply our method to the calculation of Fermi surfaces and the Elliott-Yafet spin-mixing parameter (EYP) in elemental non-magnetic metals. We find a surprisingly high anisotropy of the EYP in uniaxial hcp crystals, which can reach gigantic values as large as $830\%$ among $5d$ metals with strong spin-orbit coupling, as opposed usually less than 1\% in $5d$ cubic crystals. Through a band structure analysis, we trace this qualitative difference back to the emergence of very anisotropic spin-flip hot loops, which are supported through non-symmorphic space group of the hcp-crystal structure. We carefully investigate the effect of an external magnetic field on the spin-flip hot loops, and estimate the stability of the EYP anisotropy. We furthermore consider $3d$ and $4d$ non-magnetic elemental metals with hcp-crystal structure, where spin-orbit coupling is much weaker compared to $5d$ metals. We find a huge increase of the EYP anisotropies, reaching a colossal value as large as 6000\% for hcp Ti. We attribute this non-intuitive behavior to a different scaling of the spin-mixing parameter with respect to the atomic number between ordinary regions and spin-flip hot regions on the Fermi surface.

The paper is organized as follows: we first shortly review in Sec.~\ref{Sec:eyaf} the basics of Elliott-Yafet theory, followed by a description of our developed Fermi-surface method within the KKR method in Sec.~\ref{Sec:Method}. The successful application of our method to various non-magnetic elemental metals and the investigation of the spin-mixing parameter is presented in Sec.~\ref{Sec:Applications}, followed by conclusions in Sec.~\ref{Sec:Conclusions}. In the Appendix we discuss the possible ways of lifting the conjugation degeneracy and their physical interpretation.

\section{The Elliott approximation to spin relaxation and its anisotropy \label{Sec:eyaf}}

In this section we give a summary of previously known theoretical concepts, in order to make the paper reasonably self-contained and to define some of the quantities used later. The summary also serves as an introduction to the discussion on the physical interpretation of different ways of lifting the conjugation degeneracy, discussed in the Appendix.

One distinguishes different microscopic mechanisms causing spin-relaxation. We focus on the Elliott-Yafet mechanism, which is the dominant one in crystals with space-inversion and time-reversal symmetry, as present in the non-magnetic elemental metals that we investigate in Sec.~\ref{Sec:Applications}. For completeness, we mention that the Elliott-Yafet theory was also applied to ferromagnets \cite{Steiauf_Faehnle.2009,Carva_Oppeneer.2011}.

In the Elliott-Yafet theory, the equilibration of an excited spin-population in a non-magnetic metal occurs due to spin-flip events during scattering, which can take place \eg off impurities or phonons. The theory is based on the effect of spin-orbit coupling on the Bloch eigenstates of the crystal Hamiltonian. According to Elliott \cite{Elliott.1954}, the Bloch states are not of pure spin character, but necessarily form superpositions of spin up and spin down, written as 
\begin{eqnarray}
\boldsymbol{\Psi}_{\vcc{k} \sqa}^{+} (\vcc{r}) &=& \left[ a_{\vcc{k}  \sqa} (\vcc{r}) ~ \ket{\uparrow}_{\sqa} + b_{\vcc{k}  \sqa} (\vcc{r}) ~ \ket{\downarrow}_{\sqa} \right] ~ \expo{i \vcc{k} \cdot \vcc{r}} ~, \label{Eq:eyaf:wavefun1} \\
\boldsymbol{\Psi}_{\vcc{k} \sqa}^{-} (\vcc{r}) &=& \left[ a_{-\vcc{k} \sqa}^{*} (\vcc{r}) ~ \ket{\downarrow}_{\sqa} - b_{-\vcc{k} \sqa}^{*} (\vcc{r}) ~ \ket{\uparrow}_{\sqa} \right] ~ \expo{i \vcc{k} \cdot \vcc{r}}~. \label{Eq:eyaf:wavefun2}
\end{eqnarray}
The first equation expresses the Bloch eigenstate $\boldsymbol{\Psi}_{\vcc{k} \sqa}^{+} (\vcc{r})$ in the spin basis $(\ket{\uparrow}_{\sqa},\ket{\downarrow}_{\sqa})$ of eigenstates of the Pauli spin operator $\pauli=(\paulix,\pauliy,\pauliz)$ along a certain spin quantization axis \sqa, \ie, eigenstates of the operator $\pauli\cdot\sqa$. Usually one chooses \sqa\ along the $z$ axis, but in the present work we will allow \sqa\ to vary, exploring the spin-relaxation anisotropy. $a_{\vcc{k} \sqa} (\vcc{r})$ and $b_{\vcc{k} \sqa} (\vcc{r})$ are the lattice-periodic parts of the Bloch function.  \teqref{Eq:eyaf:wavefun2} follows from \teqref{Eq:eyaf:wavefun1} in the presence of combined time-reversal (absence of magnetic fields) and space-inversion symmetry, and the degeneracy in $E$ and $\vcc{k}$ implied by Eqs.~(\ref{Eq:eyaf:wavefun1},\ref{Eq:eyaf:wavefun2}) is called \emph{conjugation degeneracy}, following Yafet \cite{Yafet.1963}. Then, $\boldsymbol{\Psi}_{\vcc{k} \sqa}^{+}$ and $\boldsymbol{\Psi}_{\vcc{k} \sqa}^{-} = P\,K\,\boldsymbol{\Psi}_{\vcc{k} \sqa}^{+}$ form a conjugate pair, where $P$ is the space-inversion and $K$ is the time-reversal operator (see Appendix).
Defining the spin expectation value of $\boldsymbol{\Psi}_{\vcc{k} \sqa}^{\pm}$ along \sqa\ as 
\begin{equation}
S^{\pm}_{\vcc{k}\sqa}:= \frac{\hbar}{2}\bra{{\Psi}_{\vcc{k} \sqa}^{\pm}} \pauli\cdot\sqa\ket{{\Psi}_{\vcc{k} \sqa}^{\pm}}, 
\label{eq:sev}
\end{equation}
we have $S^{-}_{\vcc{k}\sqa}=-S^{+}_{\vcc{k}\sqa}$.
Due to the conjugation degeneracy, the crystal Hamiltonian together with the translational operator can only define the subspace spanned by $\boldsymbol{\Psi}_{\vcc{k} \sqa}^{\pm}$, and an additional condition is used to uniquely determine  each state: one demands that  $S^{+}_{\vcc{k}\sqa}$ is maximal (this choice physically motivated but not the only possible one; see the Appendix) and in this way $a_{\vcc{k}  \sqa}$ and $b_{\vcc{k}  \sqa}$ are uniquely defined up to an arbitrary global phase.

Following Fabian \cite{Fabian:HotSpots:1998}, we define the volume integral over the coefficients as $\bsqks = \int { \lvert b_{\vcc{k}  \sqa} (\vcc{r}) \rvert^2 \mathrm{d}\vcc{r}}$ (and equivalently for $\asqks$). For the coefficients we have by normalization $\asqks + \bsqks = 1$ and by definition $\asqks \geq \bsqks$, thus the \emph{spin-mixing parameter} $\bsqks$ determines the amount of spin-down character ``mixed'' in a predominantly spin-up state.
Obviously,
\begin{equation}
   S^{+}_{\vcc{k}\sqa} = \frac{\hbar}{2} \left( 1 - 2 \bsqks \right)~.
\end{equation}

In Elliott's equation for the spin-relaxation time, the Fermi-surface (FS) averaged spin-mixing, or Elliott-Yafet, parameter enters. It is given by
\begin{equation}
  \bsq = \frac{1}{n(E_\mathrm{F})} ~ \frac{1}{\hbar} \, \int_\mathrm{FS}{  \frac{  \bsqks   }{ \lvert\vcc{v}_\mathrm{F}(\vcc{k})\rvert } ~ \td{S} }~, \label{integral}
\end{equation}
where $\vcc{v}_\mathrm{F}(\vcc{k})$ is the Fermi velocity. The normalization by the density of states at the Fermi level, $n(E_F) = (1/\hbar) \, \int_\mathrm{FS}{ \lvert \vcc{v}_\mathrm{F}(\vcc{k}) \rvert^{-1} ~ \td{S} }$, ensures that $ 0\leq \bsq \leq 0.5$. Since the value of $\bsq$ depends on the choice of \sqa\ (as has been shown before for several types of systems \cite{Zimmermann:Spinrelaxation:2012,Mokrousov:Review:2013,Long:HotSpots:2013,Long:W001:2013,Long:SpinRelaxHall:2014} and as we discuss in the present paper), we may introduce the anisotropy that $\bsq$ shows with respect to all possible choices of \sqa:
\begin{equation}
  \mathcal{A}[b^2] = \frac{ \mathrm{max}_{\sqa} \,\bsq - \mathrm{min}_{\sqa} \,\bsq}{ \mathrm{min}_{\sqa}\, \bsq}.
\label{eq:anisotropy}
\end{equation}
The anisotropy concept can be summarized like this: if the functions $a_{\vcc{k}}$ and $b_{\vcc{k}}$ are first chosen to maximize $S^{+}_{\vcc{k}}$ along \sqa, and then are chosen to maximize $S^{+}_{\vcc{k}}$ along another axis $\sqa'$, then the two values of $S^{+}_{\vcc{k}}$ will be in general different unless \sqa\ and $\sqa'$ are symmetry-related by the crystal structure.

The central ansatz of the Elliott-Yafet theory is that electrons beyond spin-equilibrium populate $\boldsymbol{\Psi}_{\vcc{k} \sqa}^{+}$ states, while scattering from $\boldsymbol{\Psi}_{\vcc{k} \sqa}^{+}$ to $\boldsymbol{\Psi}_{\vcc{k}' \sqa}^{-}$ produces spin flips, eventually restoring equilibrium.
The practical importance of $\bsq$ and $\mathcal{A}[b^2]$ becomes clear when considering the Elliott approximation \cite{Elliott.1954} that relates $\bsq$ to the ratio of the spin-relaxation time $T_1$ and momentum relaxation time $\tau$,
\begin{equation}
 \frac{\tau}{T_1} = p \, \bsq
\end{equation}
with a proportionality constant $p$ of order one. Here, $1/T_1$ represents the spin-flip transition rate averaged over the Fermi surface, while $1/\tau$  represents the total (spin-conserving plus spin-flip) decay rate averaged over the Fermi surface, both due to scattering. $1/\tau$ shows no anisotropy with respect to \sqa. Thus, through the anisotropy of $\bsq$, an anisotropy of spin-relaxation time is induced \cite{Zimmermann:Spinrelaxation:2012}, corresponding to different values of $T_1$ depending on the spin-direction of the injected electrons in a material:
\begin{equation}
  \mathcal{A}[T_1] = \frac{ \mathrm{max}_{\sqa} \,T_1(\sqa) - \mathrm{min}_{\sqa}\, T_1(\sqa)}{ \mathrm{min}_{\sqa}\, T_1(\sqa)}.
\end{equation} 
Interestingly, $\mathcal{A}[b^2]$ stems from the band structure alone, because no specific assumptions on the scattering potential are made in the derivation of the Elliott approximation. Explicit calculations of self-adatom impurities on metallic films, where all details of the scattering potential have been included \cite{Long:SpinRelaxHall:2014} have shown that the anisotropy $\mathcal{A}[T_1]$ is in many cases qualitatively well described by the lowest-order approximation, $\mathcal{A}[b^2]$, alone.

The spin-mixing parameter is usually small ($\bsqks \ll 0.5$), but may reach the maximal value of $0.5$ for special points in the band structure (called \emph{spin-flip hot-spots}) \cite{Fabian:HotSpots:1998}. For a deeper analysis of the origin of the spin-flip hot spots, it is insightful to divide the spin-orbit operator into a spin-conserving part, $\xi(LS_\parallel)$, and a spin-flip part, $\xi(LS^{\uparrow \downarrow})$, given respectively by the first and second part on the r.h.s.\ of the following expression:
\begin{equation}
  \xi \vcc{L} \cdot \vcc{S} = \xi L_{\sqa} S_{\sqa} + \xi \left( L_{\sqa}^{+} S_{\sqa}^{-} + L_{\sqa}^{-} S_{\sqa}^{+} \right)/2\,. \label{Eq:spinorbit}
\end{equation}
Here, $\xi(r)$ is the spin-orbit coupling strength, $\vcc{L}$ and $\vcc{S}=\frac{\hbar}{2}\pauli$ are respectively the orbital and spin angular momentum operators, $L_{\sqa} = \vcc{L} \cdot \sqa$, $S_{\sqa} = \vcc{S} \cdot \sqa$, and $L_{\sqa}^{\pm}$ and $S_{\sqa}^{\pm}$ are the corresponding raising and lowering operators for angular momentum and spin in the reference frame specified by $\sqa$. It is clear that the dot product $\vcc{L} \cdot \vcc{S}$ is independent of $\sqa$, leaving the eigenenergies of the Hamiltonian invariant. However, the matrix elements of the spin-conserving and spin-flip parts, respectively, depend on the choice of the SQA. Evidently, only the spin-flip part of SOC causes a spin-mixing of the Bloch states as it has off-diagonal components as a matrix in spin space, and the spin-conserving part is a diagonal matrix in spin space.

\section{Method\label{Sec:Method}}

\subsection{KKR band structure formalism}

\subsubsection{Band structure}

To calculate the electronic band structure of crystals, we employ density functional theory (DFT) in the local density approximation (LDA). Before we turn to the details of the tetrahedron method, we recall the basic equations for calculating the electronic structure within the Korringa-Kohn-Rostoker (KKR) Green function method. The KKR secular equation reads
\begin{equation}
  \mat{M}(\vcc{k},E) ~ \vc{c}(\vcc{k},E) = 0 \,,
  \label{KKR:secularequation}
\end{equation}
where the KKR-matrix $\mat{M}(\vcc{k},E) = \mat{1} - \mat{g}(\vcc{k},E) \, \mat{t}(E)$ contains the Fourier-transformed structure constants of free space \begin{equation}
  g_{\Lambda \Lambda'}^{\mu \mu'}(\vcc{k},E) = \sum_{n,n'}{ \expo{\imag \vcc{k} \cdot (\vcc{R}_n-\vcc{R}_{n'})} \, g_{\Lambda \Lambda'}^{\mu \mu'}(\vcc{R}_n-\vcc{R}_{n'}, E)} \label{KKR:FTstructureConstants}
\end{equation}
and atomic transition matrix $\mat{t}(E) = \lbrace \delta_{\mu \mu'} \, t^{\mu}_{\Lambda \Lambda'} \rbrace$. These matrices and the vectors $\vc{c}(\vcc{k},E) = \lbrace c_{\Lambda}^{\mu} \rbrace$ depend on the combined angular momentum and spin index $\Lambda = (\ell, m, \sigma)$ and $\mu$ labels the atoms in the unit cell, and $\vcc{R}_n$ is a lattice vector. We use the \emph{screened} KKR formalism \cite{screened_KKR}, where the structure constants of free space are replaced by the ones of a reference system of repulsive muffin-tin potentials with transition matrices $\mat{t}^\mathrm{ref}$. The resulting structural Green function, $g_{\Lambda \Lambda'}^{\mathrm{ref},\mu \mu'}(\vcc{R}_n-\vcc{R}_{n'},E)$ decays rapidly with distance. Then the secular equation takes the form
\begin{equation}
    \mat{\tilde{M}}(\vcc{k},E) ~ \vc{c}(\vcc{k},E) = 0 \,,
      \label{KKR:secularequation_ref}
\end{equation}
where $\mat{\tilde{M}}(\vcc{k},E) = \mat{1} - \mat{g}^{\mathrm{ref}}(\vcc{k},E) \, \Delta \mat{t}(E)$, and $\Delta \mat{t}(E) = \mat{t}(E) - \mat{t}^\mathrm{ref}(E)$. The two KKR-matrices in Eqs.~(\ref{KKR:secularequation},\ref{KKR:secularequation_ref}) are connected to each other via
\begin{equation}
  \mat{\tilde{M}} = \left( 1-\mat{g} \, \mat{t}^\mathrm{ref} \right)^{\!\!-1} \mat{M}~.
\end{equation}
Hence, as long as the first term on the r.h.s.\ does not become singular for the energy-range of interest (which is typically true for energies as large as 30~eV above the Fermi level \cite{screened_KKR}), the roots of the two secular equations and the coefficient vectors are identical.

The vector $\vc{c}$ contains the expansion coefficients of the wavefunction in terms of the regular scattering solutions, $R^{\mu}_{\Lambda' \Lambda}(r,E)$, of the radial Schr\"odinger equation off the potentials at sites $\vcc{\tau}_{\mu}$,
\begin{eqnarray}
  \boldsymbol{\Psi}_{\vcc{k}j}(\vcc{r} + \vcc{\tau}_{\mu}) = \sum_{\Lambda'}{ Y_{L'}(\vcn{r}) \, \boldsymbol{\chi}^{\sigma'} \sum_{\Lambda}{ R^{\mu}_{\Lambda' \Lambda}(r,E) \, c^{\mu}_{\Lambda,j}(\vcc{k},E) }}\,. \nonumber \\ 
 \label{KKR:wavefunction}
\end{eqnarray}
Here, the real-space vector $\vcc{r}$ is restricted to the atomic cell around the atom position $\vcc{\tau}_{\mu}$, the spinors $\boldsymbol{\chi}^{\uparrow}=\SmallMatrix{1\\0}$ and $\boldsymbol{\chi}^{\downarrow}=\SmallMatrix{0\\1}$ define a basis in spin-space, and $r$ and $\vcn{r}$ denote the modulus and direction, respectively, of $\vcc{r}$. $Y_{L}$ denotes spherical harmonics of angular momentum $L=(\ell,m)$. The index $j$ labels possible degeneracies, for example as it is the case for non-magnetic hosts with space-inversion symmetry (see Sec.~\ref{Sec:eyaf} and the Appendix). In case of a degeneracy, the eigenvectors $\vc{c}_j$ corresponding to different orthonormal eigenfunctions $\boldsymbol{\Psi}_{\vcc{k}j}$ are not themselves orthonormal; an orthogonalization of the wavefunctions is then needed, taking into account the full form \eqref{KKR:wavefunction}.

If one considers instead of the Schr\"odinger equation the scalar-relativistic equation or Dirac equation, then the regular solutions $R(r,E)$ in Eq.~\eqref{KKR:wavefunction} have a large and small component and the wavefunction turns into a four-component spinor. The regular solutions obey a Lippmann-Schwinger equation that are solved non-iteratively as described in Ref.~\cite{Bauer.phd}.

For a given pair $(\vcc{k},E)$, \teqref{KKR:secularequation} only yields a non-zero coefficient vector $\vc{c}$ (and thus a non-vanishing wavefunction via \teqref{KKR:wavefunction}), if the corresponding KKR matrix is singular. These pairs define the band structure $E(\vcc{k})$ of the crystal.
We stress that the band structure is defined implicitly via the KKR secular equation \eqref{KKR:secularequation}, in contrast to the explicit calculation of $E(\vcc{k})$ via a diagonalization of the Hamiltonian in basis-set based methods.

\subsubsection{Spin expectation-value} \label{sec:KKR:spinvalue}

Knowing the wave function $\boldsymbol{\Psi}_{\vcc{k}j}(\vcc{r} + \vcc{\tau}_{\mu})$, the spin-expectation value $S_{\vcc{k}j}$ of a state is given by
\begin{equation}
  {S}_{\vcc{k}j} = \frac{\hbar}{2} \, \sum_{\mu}\int_{V^{\mu}} \mathrm{d}^3 r ~ \boldsymbol{\Psi}^{\dagger}_{\vcc{k}j}(\vcc{r} + \vcc{\tau}_{\mu})  ~ \left( \pauli \cdot \sqa \right)~ \boldsymbol{\Psi}_{\vcc{k}j}(\vcc{r} + \vcc{\tau}_{\mu})~.
  \label{KKR:spinexpval:wavefun}
\end{equation}
In a ferromagnet the SQA is given by the direction of the magnetization. The case of a non-magnetic and space-inversion symmetric host is described in Sec.~\ref{Sec:eyaf} and in the Appendix.
By inserting the expansion \teqref{KKR:wavefunction}, \teqref{KKR:spinexpval:wavefun} can be rewritten as
\begin{equation}
  {S}_{\vcc{k}j} = \frac{\hbar}{2} \, \vc{c}^\dagger_{j}(\vcc{k}) ~ \left( \vcc{\mat{\Sigma}} \cdot \sqa \right) ~ \vc{c}_{j}(\vcc{k})~,
  \label{KKR:spinexpval:coeff}
\end{equation}
where each component of $\vcc{\mat{\Sigma}} = ( \mat{\Sigma}_x, \mat{\Sigma}_y, \mat{\Sigma}_z)^\mathrm{T}$ contains the corresponding $(2\times2)$-Pauli matrix $\sigma^\mathrm{P}_i$ ($i=x,y,z$) and the regular solutions $R^{\mu}_{\Lambda' \Lambda}$,
\begin{eqnarray}
  \Sigma^{\mu,i}_{\Lambda \Lambda'}(E) &=& \sum_{L_1,\Lambda_2,\Lambda_3} C_{L_1,L_2,L_3} \int \td r ~ \Theta^{\mu}_{L_1}(r) \times \\
   && \times  \left[ R^{\mu}_{\Lambda_2 \Lambda}(r;E) \right]^{*} ~ \left( \sigma^\mathrm{P}_{i} \right)^{\sigma_2 \sigma_3} ~ R^{\mu}_{\Lambda_3 \Lambda'}(r;E)  \,. \nonumber
\end{eqnarray}
Here, $C_{L_1,L_2,L_3}$ are the Gaunt coefficients (integrated products of three spherical harmonics) and $\Theta^{\mu}_{L_1}(r)$ are the shape functions \cite{Stefanou1990231,Stefanou_Zeller_91} confining the integration to the volume of the atomic cell $\mu$.
In this way, the spin-matrix elements between Bloch states are expressed by a $\vcc{k}$-independent (but energy-dependent) matrix $\vcc{\mat{\Sigma}}$ and $\vcc{k}$-dependent eigenvectors $\vc{c}_{\nu}$. Algebraic manipulations involving linear combinations of $\boldsymbol{\Psi}_{\nu}$ (see, \eg, the Appendix) are thus merely transformed to manipulations containing the eigenvectors $\vc{c}_{\nu}$.

\subsection{Fermi-surface calculation\label{sec:KKR:fermisurface}}

In this section, we present details of our implementation of solving the KKR secular equation \eqref{KKR:secularequation_ref} to find the Fermi surface of a metal.

\subsubsection{Tetrahedron method}

To calculate the Fermi surface in practice, we fix the energy $E=E_\mathrm{F}$ in the secular equation \eqref{KKR:secularequation_ref} and drop it in our notation for simplicity. We scan the reciprocal space for a singular KKR matrix by reformulating the secular equation as an eigenvalue problem,
\begin{equation}
  \mat{M}(\vcc{k}) ~ \vc{c}_{j}(\vcc{k}) = \lambda_{j}(\vcc{k}) ~ \vc{c}_{j}(\vcc{k}) ~.
  \label{KKR:secularequation:eigenvalue}
\end{equation}
The size $N=2 \, N_\mathrm{at} \, (\ell_\mathrm{max}+1)^2$ of the matrix $\mat{M}$ is determined by the angular momentum cutoff $\ell_\mathrm{max}$, the number of atoms in the unit cell $N_\mathrm{at}$ and a factor 2 for spin. Evidently, for each matrix $\mat{M}$, also $N$ eigenvalues and eigenvectors exist, which are labeled by $j$. A Fermi-surface point $\vcc{k}_\mathrm{F}$ is found if at least one of these eigenvalues vanishes, $\lambda_{o}(\vcc{k}_\mathrm{F})=0$, and the corresponding eigenvector $\vc{c}_{o}(\vcc{k}_\mathrm{F})$ is proportional to the coefficient vector from \teqref{KKR:secularequation} [a correct normalization of the wavefunction in \teqref{KKR:wavefunction} has to be ensured]. If the corresponding state is $n$-fold degenerate, also $n$ eigenvalues vanish at the same band structure point $(\vcc{k}_\mathrm{F},E_\mathrm{F})$.

\begin{figure}[t!]
  \includegraphics[width=80mm]{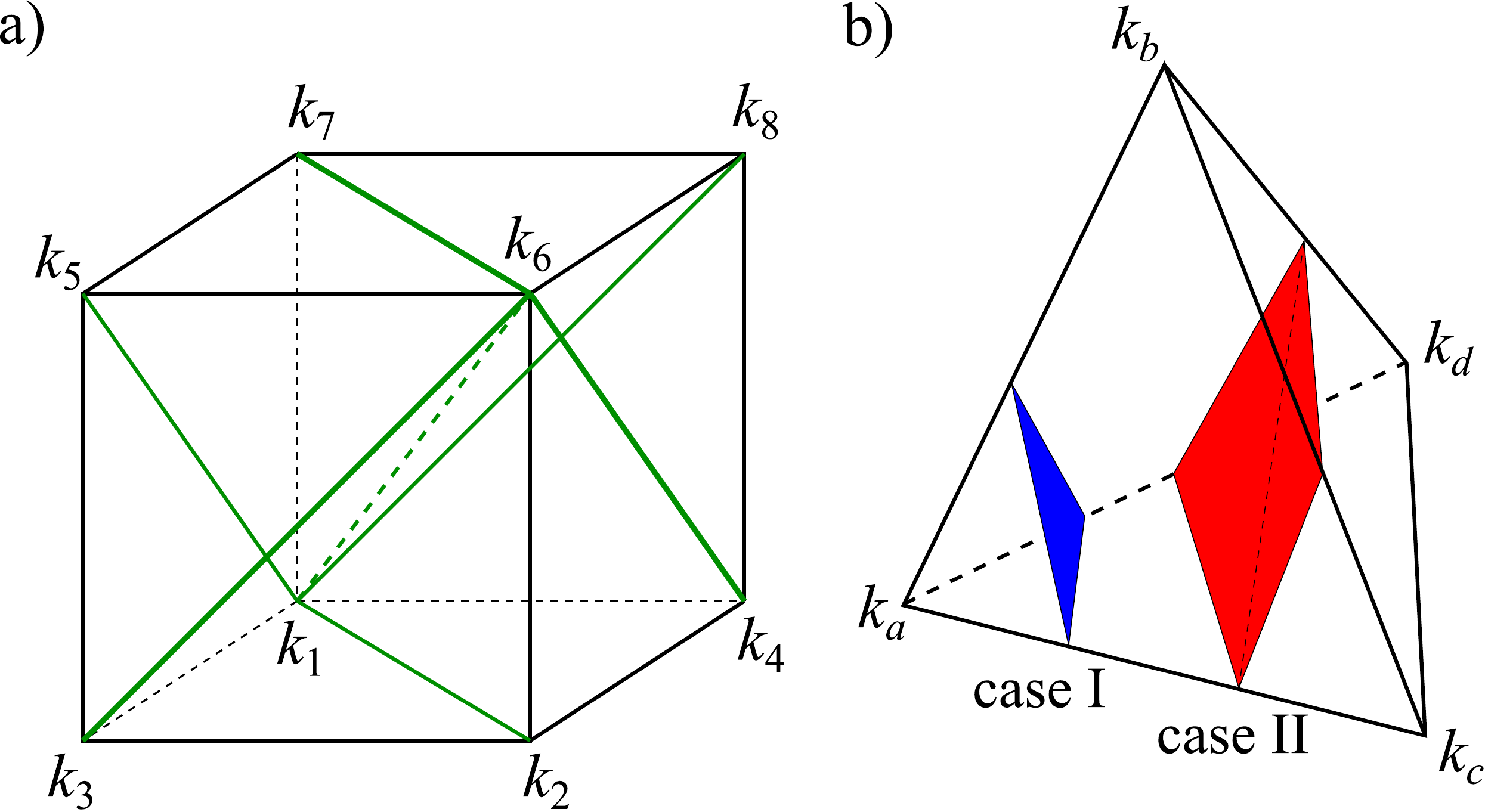}
  \caption{(color online) Left: cuboid in $k$-space for the special case of a cubic unit cell and division into 6 tetrahedra, each one defined by four vertices: (1236), (1356), (1576), (1246), (1486) and (1876). All 6 tetrahedra have a common edge, which is a space diagonal (from $k_1$ to $k_6$). Right: intersection area of a tetrahedron with the Fermi surface. Within linear interpolation, only plane objects (triangles in case I or quadrangles in case II) can occur.}
  \label{Fig:tetra}
\end{figure}

To find the points with $\lambda(\vcc{k})=0$, we divide the reciprocal space into a set of space-filling tetrahedra. First, a regular rectangular grid is created with 8 neighboring grid points forming a cuboid. Then, each cuboid is further divided into 6 tetrahedra (\cf Fig.~\ref{Fig:tetra}a). As a next step, the roots of the eigenvalues $\lambda_i$ are searched for along the edges of a tetrahedron. Finally, the intersection area of the Fermi surface with the tetrahedron is determined. Three cases can be distinguished \cite{Bloechl:tetramethod}: a triangle (case I) or a quadrangle (case II, \cf Fig.~\ref{Fig:tetra}b) or no intersection area (case III). For sake of simplicity in a computer code, a quadrangle can be decomposed into two triangles and no distinction between the cases I and II has to be made in subsequent parts of the code.

We want to stress some computational aspects:
\begin{enumerate}

\item{The KKR matrix is in general non-hermitian and as a result the eigenvalues are complex numbers. The real and imaginary part of $\lambda_{o}$ do not necessarily vanish at the exact same $k$ point due to finite numerical cutoff parameters. We determine the root such, that the imaginary part vanishes and check whether also the real part is reasonably small. We usually achieve $\Im \lambda_{o} \sim 10^{-12}$ and $\Re \lambda_{o} \sim 10^{-5}$.}

\item{To find a root along an edge, we compute the eigenvalues at the start and end points of the edge and interpolate linearly in between. By doing so, we find an approximate Fermi-surface point where the linearly interpolated eigenvalue vanishes. However, the true intersection point of the Fermi surface with the edge will be somewhat different, and we refine the approximate $k$-point by applying a nested intervals method (false position method, \cf Ref.~\cite{numerical_recipies}). Usually, only three to five iterations are needed to ensure $\Im \lambda_{o}=0$ up to the precision stated above.}

\item{The order of the (complex valued) eigenvalues depends on the computer routine which is used to diagonalize the KKR matrix. Thus, when comparing the eigenvalues at two different $\vcc{k}$ points, $\lambda_i(\vcc{k}_1)$ and $\lambda_j(\vcc{k}_2)$, the connectivity ($i \leftrightarrow j$) that should correspond to the continuity of $\lambda_i(\vcc{k})$ is not known a-priori. As a result, it is not possible to interpolate the eigenvalues between discrete $\vcc{k}$ points. To resolve this issue, we use the fact that the coefficient vectors are (nearly) orthogonal to each other if they belong to different bands and the two $\vcc{k}$ points are not too far away from each other. We calculate the pairwise projections of these coefficient vectors, $p_{ij} = \bar{\vc{c}}_{i}(\vcc{k}_1)  \cdot \vc{c}_{j}(\vcc{k}_2)$. Here, $\bar{\vc{c}}_{i}(\vcc{k}_1)$ denotes a left eigenvector of $\mat{M}(\vcc{k}_1)$, \ie
\begin{equation}
  \bar{\vc{c}}_{i}(\vcc{k}_1) ~ \mat{M}(\vcc{k}_1)  = \lambda_{i}(\vcc{k}_1) ~ \bar{\vc{c}}_{i}(\vcc{k}_1) ~.
  \label{KKR:secularequation:lefteigenvector}
\end{equation}
For a selected $i$, we find $p_{ij} \approx 1$ only for one $j \in \lbrace 1, \ldots, N \rbrace$, which determines the connection between the eigenvalues at $\vcc{k}_1$ and $\vcc{k}_2$.

If the system exhibits the aforementioned conjugation-degeneracy, each state is two-fold degenerate and, as a consequence, always two eigenvalues (say $\lambda_i$ and $\lambda_{i+1}$) are the same. Then, we only treat one of the two degenerate eigenvalues at $\vcc{k}_{1}$ (\ie all $\lambda_{2i}$ for $i\in \lbrace 1, \ldots, N/2 \rbrace$), but still calculate the projections $p_{ij}$ for all $j \in \lbrace 1, \ldots, N\rbrace$ at $\vcc{k}_{2}$. In the worst case, $\max_{j} p_{ij} \approx 0.5$ for the two $j$s that belong to the conjugation-degenerate pair and very small for the other $j$s.

We highlight that due to the correct connectivity of the eigenvalues, the method is capable of calculating crossings of Fermi-surface sheets correctly.}
\end{enumerate}

\subsubsection{Visualization set}
At the end, the whole Fermi surface is represented as a collection of triangles from all tetrahedra. Evidently, neighboring tetrahedra share intersection points, and thus the number of \emph{distinct} $\vcc{k}$ points is much smaller than three times the number of triangles (typically by a factor of five to six for bulk crystals).
This set can still be utilized to visualize the Fermi surface and calculated properties on it, hence we call it \emph{visualization set}.

\subsubsection{Integration set\label{sec:KKR:integration_set}}
It may be required to further reduce the number of $\vcc{k}$ points on the Fermi surface. This is especially the case, when the quantity to be calculated is a function of two or more $\vcc{k}$ points. A prominent example is the scattering rate $P_{\vcc{k} \vcc{k'}}$ needed in the calculation of electron transport properties or spin and momentum relaxation times. We achieve a further reduction of the number of $\vcc{k}$ points by first merging all triangles that originate from a cuboid (remember that a regular rectangular mesh underlies the tetrahedra) into a set of triangles. Then, this whole set is represented by a single $\vcc{k}$ point, which is chosen to be the closest one to the center of the cuboid. A weight $S_\vcc{k}$ of this representative $\vcc{k}$ point is given by the total area of the triangles in this set. If more than one Fermi-surface sheet intersects the cuboid, each sheet is represented by its own $\vcc{k}$ point and weight. All representative $\vcc{k}$ points form the so-called \emph{integration set}. It is not possible anymore to visualize this set, because the information about the explicit form of the Fermi surface is lost by concatenating it into the weights $S_\vcc{k}$. However, the integration set is well suited to perform accurate Fermi-surface integrals.

\subsection{Fermi velocity\label{sec:KKR:fermivelocity}}

The Fermi velocity $\vcc{v}_\mathrm{F}(\vcc{k}) = (\partial E/ \partial \vcc{k}) \vert_{E=E_\mathrm{F}}$, where $E$ is the band energy under consideration, is often required for the evaluation of Fermi-surface integrals since it appears in the integration weight, $\td{S}/\abs{\vcc{v}_\mathrm{F}(\vcc{k})}$. 

In analogy to Gradhand \etal \cite{Gradhand.Berry.2011}, the Fermi velocity is calculated via the derivative of the KKR-matrix eigenvalues $\lambda_{o}$ with respect to $\vcc{k}$ and $E$,
\begin{equation}
  v_\mathrm{F}^i(\vcc{k}) = - \frac{ \partial \lambda_{o} / \partial k^i}{ \partial \lambda_{o} / \partial E } \bigg|_{\vcc{k}=\vcc{k}_\mathrm{F}, E=E_\mathrm{F}}\,,
\end{equation}
where the superscript $i \in \lbrace x,y,z\rbrace$ denotes the Cartesian component of a vector. The derivatives are calculated from finite differences through a two-point rule.
For the energy derivative, we diagonalize $\mat{M}(\vcc{k}_\mathrm{F},E_\mathrm{F})$ and $\mat{M}(\vcc{k}_\mathrm{F},E_\mathrm{F} \pm \delta E)$. A proper connection between the vanishing eigenvalue $\lambda_{o}$ at $E_\mathrm{F}$ and the corresponding eigenvalues at $E_\mathrm{F} \pm \delta E$ has to be ensured as explained in the previous Sec.~\ref{sec:KKR:fermisurface}. The derivative with respect to $\vcc{k}$ is done analogously between the points $\vcc{k}$ and $\vcc{k} \pm \delta k \cdot \vcn{e}^i$, where $\vcn{e}^i \in \lbrace \vcn{e}^x, \vcn{e}^y, \vcn{e}^z \rbrace$ is a unit vector. The full Fermi-velocity vector thus can be obtained. We usually choose $\delta k \approx 10^{-5}~\frac{2\pi}{a}$ and $\delta E \approx 10^{-5}$~Ryd.

\subsection{Fermi-surface integrals}

Next we want to consider integrals of the form
\begin{equation}
  A = \int\limits_\mathrm{FS}{\frac{\td{S}}{\abs{\vcc{v}_\mathrm{F}(\vcc{k})}} ~ f(\vcc{k})} ~. \label{eq:FSintegral}
\end{equation}
The division by the Fermi velocity renormalizes the infinitesimal area $\td{S}$ according to the density of states of this particular band-structure point.

For the \emph{visualization set}, the Fermi surface is represented in terms of a finite number of triangles, and the Fermi-surface integral turns into a finite sum over all these triangles, \ie $A = \sum_{t} {A_t}$. We approximate the contribution of a triangle, $A_t$, by taking the values of the function of interest on the three corner points of the triangle (\ie $f(\vcc{k}_i)$ with $i=1,2,3$) and interpolate linearly between them. The result takes the simple form
\begin{equation}
  A_t = \frac{S_t}{3}\sum_{i=1}^3\frac{f(\vcc{k}_i)}{ \abs{\vcc{v}_\mathrm{F}(\vcc{k}_i)}},
\end{equation}
where just the mean average of the integrand at the three corner points enters and $S_t$ is the area of the triangle.

For the \emph{integration set}, the integral just turns into
\begin{equation}
  A = \sum_{\vcc{k}}{ S_{\vcc{k}} ~ \frac{f(\vcc{k})}{\abs{\vcc{v}_\mathrm{F}(\vcc{k})}} } ~,
\end{equation}
where the sum is over all representative $\vcc{k}$-points in the integration set and $S_{\vcc{k}}$ are their weights (see Sec.~\ref{sec:KKR:integration_set}).

\section{Application \label{Sec:Applications}}

We apply our method to the calculation of the Fermi surfaces and the Elliott-Yafet parameter (EYP) for various metals from density functional theory (DFT) in the local spin-density approximation (LSDA) using the parametrization of Vosko, Wilk and Nusair \cite{LDA_VWN}, employing the Korringa-Kohn-Rostoker method as explained in the previous sections. We choose the experimental crystal structures with lattice constants as given in Tab.~\ref{table:EYP_all}.

The computational scheme which was used can be divided into two sets: (i) for $5d$ and $6sp$ elements, a self-consistent potential was obtained solving the fully-relativistic Dirac equation. In the final step for the determination of the Fermi surface, $\vcc{v}_\mathrm{F}$ and $\bsqks$, this potential was used to construct the scalar-relativistic equation plus the spin-orbit coupling term added in its Pauli-form (SRA+SOC) \footnote{At the time when calculations for $5d$ and $6sp$ metals were performed, the inclusion of the full potential and SOC in the self-consistency cycle in our code were not yet possible. Since for bulk metals the atomic sphere approximation (ASA) of spherically symmetric potentials is a good approximation for the band structure, we performed the calculation of $5d$ and $6sp$ metals in ASA where the Dirac equation was implemented \cite{SPR-TB-KKR}.}. The atomic sphere approximation (\ie neglecting non-spherical terms in the potential) was used. For the expansion of wave-functions, an angular-momentum cutoff of $\ell_\mathrm{max}=4$ was used. (ii) In contrast, for the lighter hcp elements ($3sp$, $3d$ and $4d$), within all steps the SRA+SOC-equation was used and the full potential was taken into account as it became available by the development of a new solver \cite{Bauer.phd}. An angular momentum cutoff $\ell_\mathrm{max}=3$ was found to be sufficient. Tests for W and Re have shown that procedures (i) and (ii) lead to the same results. \footnote{The values for $\bsq$ might vary, but the order of magnitude and qualitative behavior, especially for the anisotropy of $\bsq$, remain the same (see Tab.~\ref{table:EYP_all} for W and Re).}

As analyzed recently by us \cite{Zimmermann:Spinrelaxation:2012, Mokrousov:Review:2013}, the EYP can exhibit a strong anisotropy when the direction of the SQA is varied with respect to the lattice of the crystal [see Eq.~\eqref{eq:anisotropy}]. The anisotropy can reach gigantic values in systems with lowered symmetry, such as uniaxial bulk crystals \cite{Zimmermann:Spinrelaxation:2012} or thin films \cite{Long:W001:2013,Long:HotSpots:2013,Long:SpinRelaxHall:2014}. To a large extent, the anisotropy stems from points on the Fermi surface, where spin-flip hot spots exist for one direction of $\sqa$, but are absent for another direction. If such a region on the Fermi surface is rather large or forms a whole line in contrast to a singular point, we talk about a spin-flip hot \emph{area} or \emph{loop} instead of a spot. To obtain these anisotropic regions, the general rules to obtain spin-flip hot-spots as formulated by Fabian \etal \cite{Fabian:HotSpots:1998} must be met: apart from conjugation degeneracy, an additional degeneracy must be present in the scalar-relativistic (\ie without SOC) band structure. These frequently occur at Brillouin zone boundaries and along high-symmetry lines, or accidentally at an arbitrary point in the BZ. Upon inclusion of SOC, a splitting occurs that, if caused by the spin-flip part of SOC, leads to a spin-flip hot spot. Such a spot becomes in addition very anisotropic, if the electronic wavefunction exhibits particular orbital character \cite{Mokrousov:Review:2013,Zimmermann:Spinrelaxation:2012}.

We first discuss the EYP and its anisotropy in the 6$sp$ metals fcc-Au, hcp-Tl and fcc-Pb and all $5d$ metals. We distinguish different directions of the SQA and exemplify the conclusions made in Ref.~\cite{Zimmermann:Spinrelaxation:2012}. We then investigate the influence of an external $B$-field on the spin-mixing parameter in these metals. Last, we present the EYP in other elemental metals with hcp crystal structure (Mg, Sc, Ti, Zn, Y, Zr, Tc, Ru and Cd).

\subsection{5$d$ and 6$sp$ metals}

\begin{table*}
	\caption{Fermi-surface averaged spin-mixing parameter $\bsq$. For, hcp and cubic crystals  two and three, respectively, high-symmetry directions of the SQA are considered, as well as the value for polycrystalline samples and the anisotropy $\mathcal{A}$ as defined in the text. The lattice parameter $a$ is given in units of Bohr radii. \label{table:EYP_all}}
	
	\begin{ruledtabular}
	\begin{tabular}{rr cc r rrrr}
		\multicolumn{9}{c}{hcp crystals}\\[1.3ex] \hline
		& & \multicolumn{2}{c}{lattice parameter} & \multicolumn{5}{c}{spin-mixing parameter} \\
		& & $a$ & $c/a$ & \multicolumn{1}{c}{$\mathcal{A}$}& $c$ axis & $ab$ plane &  & Polycrystal  \\[1.1ex] 
		\cmidrule[0.4pt]( lr{0.4em}){3-4}%
		\cmidrule[0.4pt]( lr{0.4em}){5-9}%
		
		3$sp$ & Mg$\phantom{^\dagger}$ & $6.06$ & $1.624$ & $2200 \%$ & $2.0  \times 10^{-5}$ & $4.63 \times 10^{-4}$ & & $3.15 \times 10^{-4} ~^\ddag$ \\[1.3ex]
		
		3$d$ & Sc$\phantom{^\dagger}$ & $6.25$ & $1.594$ & $1250 \%$ & $8.5  \times 10^{-5}$ & $1.16 \times 10^{-3}$ & & $8.02 \times 10^{-4} ~^\ddag$ \\
		& Ti$\phantom{^\dagger}$ & $5.58$ & $1.588$ & $6000 \%$ & $1.77 \times 10^{-4}$ & $1.09 \times 10^{-2}$ & & $7.33 \times 10^{-3} ~^\ddag$ \\
		& Zn$\phantom{^\dagger}$ & $5.03$ & $1.856$ &  $435 \%$ & $2.59 \times 10^{-4}$ & $1.39 \times 10^{-3}$ & & $1.01 \times 10^{-3} ~^\ddag$ \\[1.3ex]
		
		4$d$ &  Y$\phantom{^\dagger}$ & $6.89$ & $1.571$ & $450 \%$  & $1.31 \times 10^{-3}$ & $7.20 \times 10^{-3}$ & & $5.24 \times 10^{-3} ~^\ddag$ \\
		& Zr$\phantom{^\dagger}$ & $6.11$ & $1.593$ & $705 \%$  & $4.51 \times 10^{-3}$ & $3.63 \times 10^{-2}$ & & $2.57 \times 10^{-2} ~^\ddag$ \\
		& Tc$\phantom{^\dagger}$ & $5.17$ & $1.604$ & $137 \%$  & $2.32 \times 10^{-2}$ & $5.51 \times 10^{-2}$ & & $4.45 \times 10^{-2} ~^\ddag$ \\
		& Ru$\phantom{^\dagger}$ & $5.11$ & $1.584$ &  $86 \%$  & $1.24 \times 10^{-2}$ & $2.31 \times 10^{-2}$ & & $1.95 \times 10^{-2} ~^\ddag$ \\
		& Cd$\phantom{^\dagger}$ & $5.63$ & $1.886$ & $202 \%$  & $1.69 \times 10^{-3}$ & $5.11 \times 10^{-3}$ & & $3.97 \times 10^{-3} ~^\ddag$ \\[1.3ex]
		
		5$d$ & La$\phantom{^\dagger}$ & $7.124$ & $1.611$ & $150 \%$ & $1.40 \times 10^{-2}$ & $3.46 \times 10^{-2}$ & & $2.62 \times 10^{-2} \phantom{~^\ddag}$ \\
		& Lu$\phantom{^\dagger}$ & $6.620$ & $1.585$ & $200 \%$ & $1.10 \times 10^{-2}$ & $3.33 \times 10^{-2}$ & & $2.53 \times 10^{-2} \phantom{~^\ddag}$  \\
		& Hf$\phantom{^\dagger}$ & $6.040$ & $1.580$ & $830 \%$ & $1.62 \times 10^{-2}$ & $1.51 \times 10^{-1}$ & & $9.55 \times 10^{-2} \phantom{~^\ddag}$  \\
		& Re$\phantom{^\dagger}$ & $5.218$ & $1.615$ & $88 \%$ & $6.42 \times 10^{-2}$ & $1.21 \times 10^{-1}$ & & $9.98 \times 10^{-2} \phantom{~^\ddag}$ \\
		& Re$^\dagger$ & $5.218$ & $1.615$ & $69 \%$ & $8.38 \times 10^{-2}$ & $1.41 \times 10^{-1}$ & & $1.22 \times 10^{-1} ~^\ddag$ \\
				
		& Os$\phantom{^\dagger}$ & $5.167$ & $1.579$ & $59 \%$ & $4.85 \times 10^{-2}$ & $7.69 \times 10^{-2}$ & & $6.66 \times 10^{-2} \phantom{~^\ddag}$ \\[1.3ex]
		
		6$sp$ & Tl$\phantom{^\dagger}$ & $6.520$ & $1.598$ & $19 \%$ & $ 5.04 \times 10^{-2}$ & $ 6.00 \times 10^{-2}$ & &  $5.61 \times 10^{-2} \phantom{~^\ddag}$ \\[3ex] 

		\multicolumn{9}{c}{cubic crystals}\\[1.3ex] \hline
		
		\multicolumn{2}{c}{} & \multicolumn{2}{c}{lattice} & \multicolumn{5}{c}{spin-mixing parameter} \\
		& & & $a$ & \multicolumn{1}{c}{$\mathcal{A}$} & $[001]$ & $[110]$ & $[111]$ & Polycrystal  \\[1.3ex] 
		\cmidrule[0.4pt]( lr{0.4em}){3-4}%
		\cmidrule[0.4pt]( lr{0.4em}){5-9}%
		
			5$d$ & Ta$\phantom{^\dagger}$ & bcc & $6.247$ & $0.2 \%$ & $1.746 \times 10^{-2}$ & $ 1.750 \times 10^{-2}$ & $1.748 \times 10^{-2}$ & $1.748 \times 10^{-2}$ \\
			&  W$\phantom{^\dagger}$ & bcc & $5.981$ & $5.7 \%$ & $6.49 \times 10^{-2}$ & $6.26 \times 10^{-2}$ & $6.14 \times 10^{-2}$ & $6.27 \times 10^{-2}$ \\
			  &  W$^\dagger$ & bcc & $5.98$ & $6.0 \%$ & $5.73 \times 10^{-2}$ & $5.52 \times 10^{-2}$ & $5.41 \times 10^{-2}$ &    \\
			& Ir$\phantom{^\dagger}$ & fcc & $7.255$ & $0.9 \%$ & $5.50	 \times 10^{-2}$ & $5.54 \times 10^{-2}$ & $5.55 \times 10^{-2}$ & $5.53 \times 10^{-2}$ \\
			& Pt$\phantom{^\dagger}$ & fcc & $7.414$ & $0.4 \%$ & $5.27 \times 10^{-2}$ & $5.26 \times 10^{-2}$ & $5.25 \times 10^{-2}$ & $5.25 \times 10^{-2}$ \\[1.3ex]
			
			6$sp$ & Au$\phantom{^\dagger}$ & fcc & $7.71$ & $0.1 \%$ & $3.248 \times 10^{-2}$ & $3.252 \times 10^{-2}$ & $3.252 \times 10^{-2}$ & $3.251 \times 10^{-2}$ \\
			& Pb$\phantom{^\dagger}$ & fcc & $9.36$ & $0.1 \%$ & $6.616 \times 10^{-2}$ & $6.609 \times 10^{-2}$ & $6.608 \times 10^{-2}$ & $6.611 \times 10^{-2}$ 	
			
	\end{tabular}
	\end{ruledtabular}
	\begin{flushleft}
	$^\dagger$ calculated with full potential\\
	$^\ddag$ approximated by $b^2_\mathrm{poly} \approx \frac{1}{3} b^2_{\sqa \parallel c} + \frac{2}{3} b^2_{\sqa \parallel ab}$\\
	\end{flushleft}
\end{table*}

The EYP of all $5d$ and some $6sp$ metals is presented in Table~\ref{table:EYP_all}. Let us first comment on the \emph{magnitude} of the EYP in those crystals: For a given SQA along the $c$-axis and the $[001]$-direction for hcp and cubic crystals, respectively, the values range between $1.10 \times 10^{-2}$ for Lu and $6.6 \times 10^{-2}$ for Pb. In this case, the large magnitude of $\bsq \approx 10^{-2}$ is mainly determined by the strong spin-orbit coupling strength. For comparison, the much lighter elements Cu and Al with weaker spin-orbit coupling have an EYP of the order of $10^{-3}$ and $10^{-5}$, respectively \cite{Fabian:HotSpots:1998,Gradhand:Spinmixing:2010}.

\begin{turnpage}
	\begin{figure*}
		\includegraphics[width=230mm]{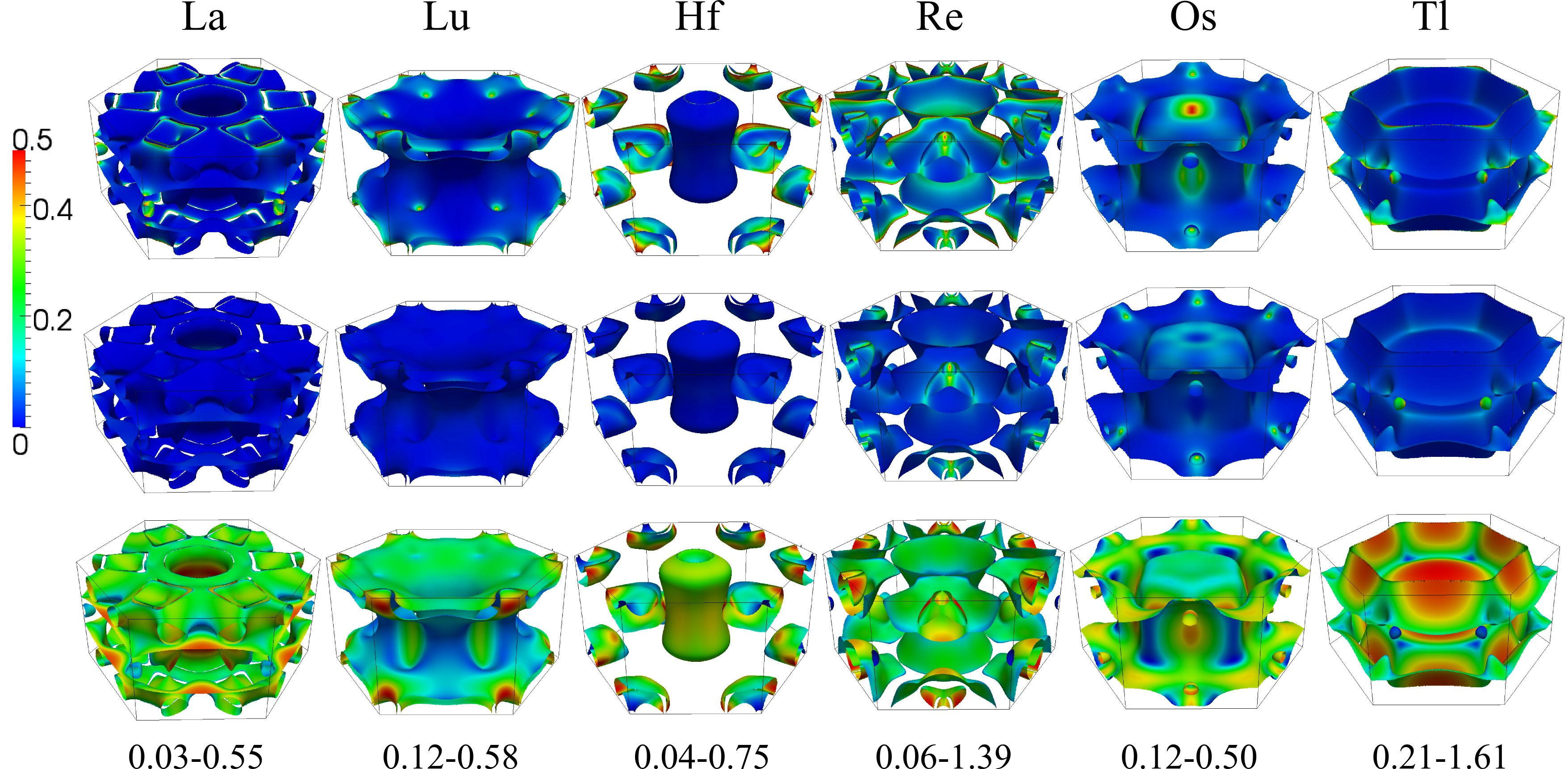}
		\caption[Spin-mixing parameter on the Fermi surface for hcp elements.]{(color online) Fermi surfaces for various $5d$ metals with hcp crystal structure. The first and second row show $\bsqks$ as color code for $\sqa$ in the $ab$-plane and along the $c$-axis, respectively. In the lower row, the absolute value of the Fermi velocity is shown. The same color legend is used as for the spin-mixing parameter, but with the limits as indicated below the plots (in atomic Rydberg units, \ie the speed of light takes the value $274.072$). \label{fig:FSgallery_hcp}}
	\end{figure*}
\end{turnpage}

The variation of $\bsq$ within the series (for fixed SQA along the $c$-axis) is determined by the details of the electronic structure. This can be seen best by comparing the distribution of the spin-mixing parameter on the Fermi-surfaces for the hcp crystals (see middle row of Fig.~\ref{fig:FSgallery_hcp}). There, the $\vcc{k}$-resolved spin-mixing parameter is shown in a color code on the Fermi-surface for all $5d$ metals with hcp crystal structure. The most important qualitative difference is the presence of spin-flip hot-spots (green to red points on the Fermi surfaces of Re, Os and Tl), which leads to an increase of the Fermi-surface averaged $\bsq$ by approximately a factor of five as compared to La, Lu and Hf (see Table \ref{table:EYP_all}). Hence, the spin-flip hot-spots have not such a dramatic impact on the averaged value as it is the case for crystals with weaker spin-orbit coupling (an increase of $\bsq$ due to spin-flip hot spots of a factor of 50 was reported for Al \cite{Fabian:HotSpots:1998}).

Let us now turn the SQA away into a different direction and investigate the \emph{anisotropy} of the EYP. As already reported in Ref.~\cite{Zimmermann:Spinrelaxation:2012}, a large anisotropy can be expected in uniaxial crystals or systems with a preferential direction, such as hcp crystals. The largest anisotropy among the $5d$ and $6sp$ elements is obtained for Hf, where the EYP increases by one order of magnitude from $1.6 \times 10^{-2}$ to $15.1 \times 10^{-2}$ when the SQA is turned from the $c$-axis to the $ab$-plane (see Table~\ref{table:EYP_all}). This corresponds to an anisotropy, defined by \teqref{eq:anisotropy}, as large as 830\%. But also the other hcp crystals exhibit a large anisotropy, where the smallest value $\mathcal{A}=19\%$ is obtained for Tl. The EYP is largest for an SQA in the $ab$-plane for all hcp crystals (see Table~\ref{table:EYP_all}).
An inspection of the Fermi-surface resolved contributions (\cf Fig.~\ref{fig:FSgallery_hcp}) reveals the emergence of large \textit{spin-flip hot areas} and \textit{hot loops} only for this direction of $\sqa$. This is the main origin of the large effect.

\begin{figure}[ht!]
  \includegraphics[width=85mm]{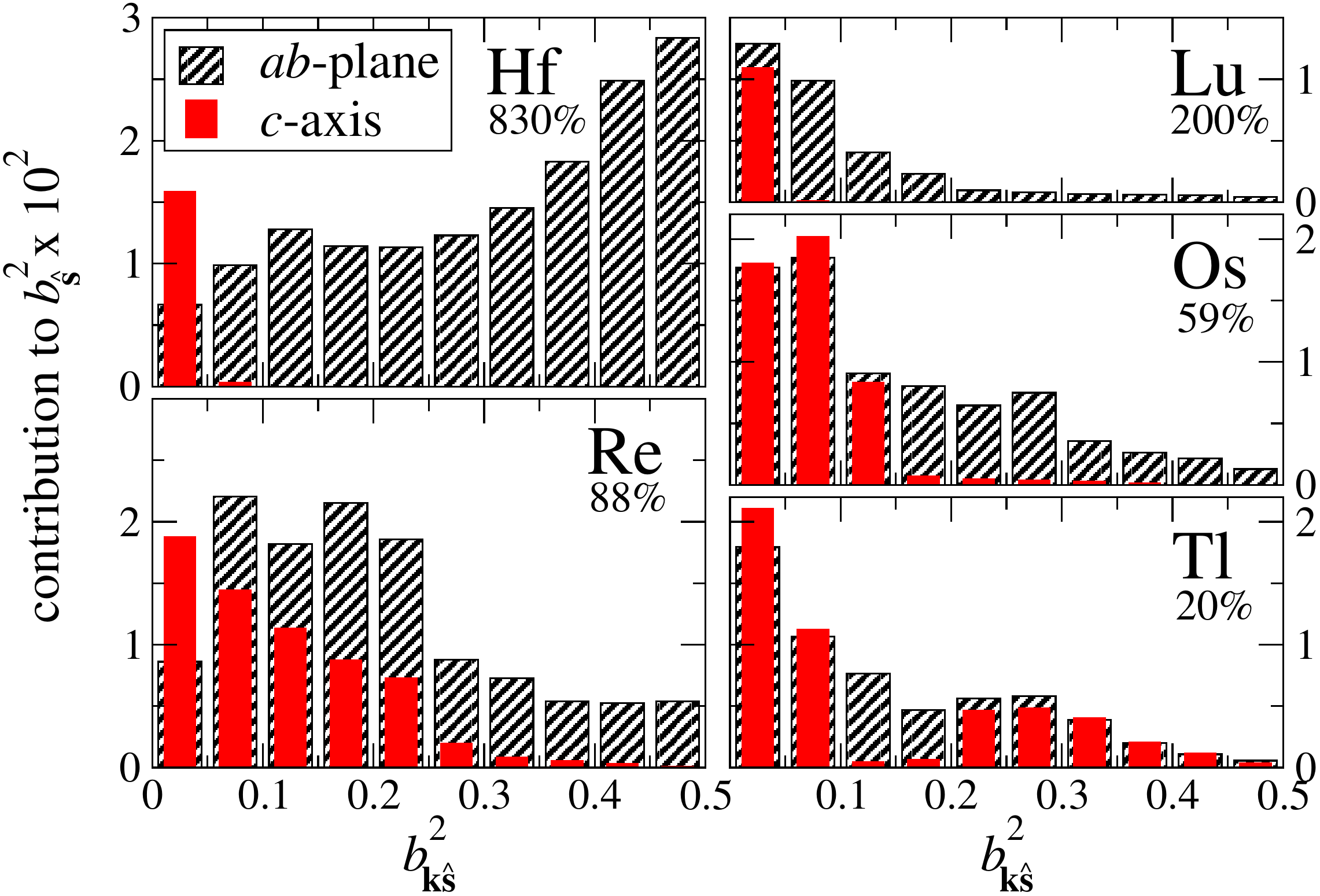}%
  \caption{(color online) Contribution to $\bsq$ according to \teqref{integral:resricted}, where the integral is restricted to regions where $\bsqks$ lies in an interval as indicated on the abscissa, for selected hcp crystals and two directions of the SQA, namely parallel to the $ab$-plane (black striped bars) and along the $c$-axis (red solid bars).}
  \label{fig:histogram}
\end{figure}

We emphasize the last point by quantifying the hot-spot contribution to the Fermi-surface averaged $\bsq$. We constrain the integral \eqref{integral} to
\begin{equation}
  \bsq = \frac{1}{n(E_\mathrm{F})} ~ \frac{1}{\hbar} \, \int_{S_i}{  \frac{  \bsqks   }{ \lvert \vcc{v}_\mathrm{F}(\vcc{k})\rvert } ~ \td{S} }~, \label{integral:resricted}
\end{equation}
where $S_i$ is the part of the Fermi surface where $\bsqks$ lies within the interval $x_i \leq \bsqks < x_{i+1}$ (with $x_i = 0, 0.05, 0.1, \ldots 0.5)$. These values form the histograms of Fig.~\ref{fig:histogram}, and in the end the sum over all parts yields the total values $\bsq$ which are presented in Table~\ref{table:EYP_all}. The giant anisotropy of Hf stems from a large interval where $0.1 \lesssim \bsqks \leq 0.5$ (compare the black striped and red solid bars in Fig.~\ref{fig:histogram}). In contrast, the interval which is relevant for the anisotropy is smaller for Os ($0.15 \lesssim \bsqks \lesssim 0.5$) and Tl ($0.1 \lesssim \bsqks \lesssim 0.2$).

\begin{turnpage}
 \begin{figure*}
  \includegraphics[width=180mm]{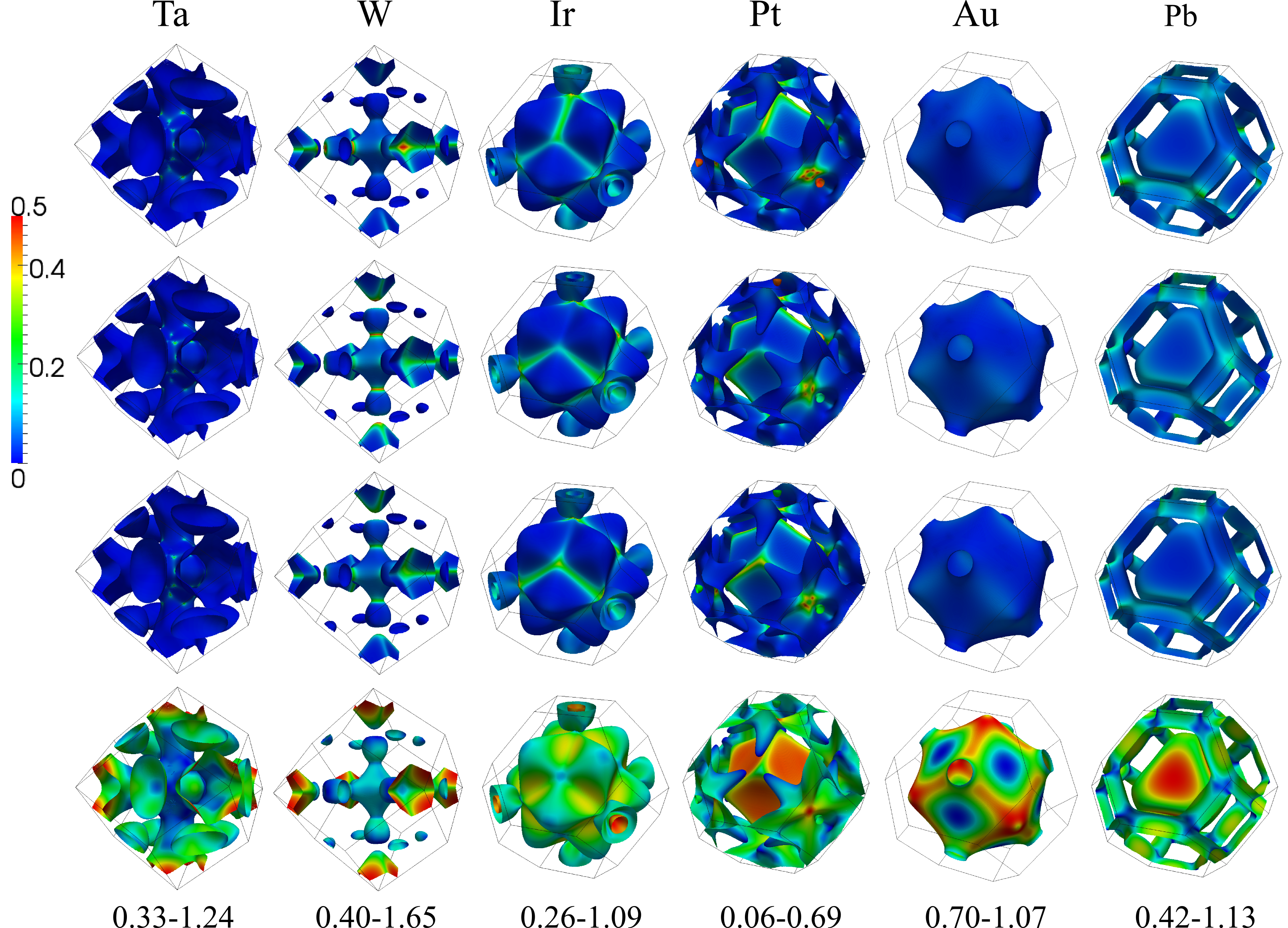}
  \caption{(color online) Fermi surfaces for various $5d$ elements with cubic crystal structure. The first, second and third row show $\bsqks$ as color code for $\sqa$ along $[001]$, $[110]$ and $[111]$, respectively. In the lower row, the absolute value of the Fermi velocity is shown. The same color legend is used as for the spin-mixing parameter, but with the limits as indicated below the plots (in atomic Rydberg units, \ie the speed of light takes the value $274.072$). \label{fig:FSgallery_cub}}
 \end{figure*}
\end{turnpage}

In Fig.~\ref{fig:FSgallery_cub}, we present the Fermi surfaces of the body- and face-centered cubic crystals and display the spin-mixing parameter on them for three high-symmetry directions of the SQA. For nearly all elements and all directions of the SQA, spin-flip hot spots --- or at least regions of strongly enhanced spin-mixing parameter --- are present.

Taking Pt as an example, for $\sqa \parallel [001]$ (which we denote as $z$-axis for simplicity), we find $\bsqks \approx 0.45$ at the four pockets in the $xy$-plane (of which only two are visible in Fig.~\ref{fig:FSgallery_cub}). Due to the cubic symmetry of the crystal, pockets with the same shape are also present along the $z$-axis, but with a low spin-mixing parameter of $\bsqks \approx 0.05$. Thus, in this case the spin-mixing parameter is high (low) if the pocket is placed perpendicular (parallel) to the SQA. This dependence is similar to the  emerging spin-flip hot areas in hcp crystals, but here we have high \emph{and} low contributions for the \emph{same} SQA. When we now change the SQA from $[001]$ to $[111]$ in Pt, all 6 pockets form an angle with the SQA. As a result, $\bsqks$ at the pockets in the $xy$-plane is reduced from $0.45$ to only approximately $0.25$. Simultaneously, the spin-mixing parameter at the two pockets along the $z$-axis increases from $0.05$ to $0.25$. The net change in the total $\bsq$ is thus strongly suppressed due to the high symmetry of the crystal, changing by merely 0.4\% (\cf Table~\ref{table:EYP_all}).

The situation is similar for the other cubic elements, e.g.\ at the handles in tungsten (see also Ref.~\cite{Zimmermann:Spinrelaxation:2012}) or at the ``hot loops'' in Ir (\cf Fig.~\ref{fig:FSgallery_cub}), and anisotropy effects at symmetry related points mainly cancel each other. As a result, the anisotropy in cubic elemental crystals is generally smaller than 1\% (see Table~\ref{table:EYP_all}), with the exception of W which exhibits a relatively strong anisotropy of about $6\%$ (\cf Table~\ref{table:EYP_all} and Ref.~\cite{Zimmermann:Spinrelaxation:2012}).

We point out, that the anisotropy is maximal for single crystals (as calculated here). In the case of polycrystals with some preferential axis orientation, the anisotropy will appear reduced, and in the case of no preferential axis it will vanish. For the latter case, the Elliott-Yafet parameter needs to be determined by averaging over all possible directions of $\sqa$,
\begin{equation}
	b^2_\mathrm{poly} = \frac{1}{4\pi} \int \td\Omega \, \bsq~,\qquad\text{with~} \sqa = \sqa(\vartheta, \varphi)~.
\end{equation}
We determined the values for  $5d$ and $6sp$ polycrystals by numerical integration over the solid angle and present them in Table~\ref{table:EYP_all}. We remark that, for hcp elements the integrand can be well approximated by a $\sin^2 \vartheta$-behavior, and the integral can be evaluated to be $b^2_\mathrm{poly} \approx \frac{1}{3} b^2_{\sqa \parallel c} + \frac{2}{3} b^2_{\sqa \parallel ab}$.

\subsection{Band-structure analysis}

In Ref.~\cite{Mokrousov:Review:2013}, some general conditions that must be met to obtain an emerging spin-flip hot spot were deduced from a simple model, in which only six $p$-states ($p_i^\sigma$, with $i=x,y,z$ and $\sigma=\uparrow,\downarrow$) were considered. Without SOC, the $p_x^\sigma$, $p_y^\sigma$ and $p_z^\sigma$-states were placed at energies $\delta/2$, $-\delta/2$ and $\Delta$, respectively, and the effect of the inclusion of SOC with strength $\xi$ was investigated. We briefly summarize the results: the largest anisotropy was obtained, if the $p_x^\sigma$ and $p_y^\sigma$-states are close in energy ($\delta \ll \xi$), and the system is highly uniaxial, $\Delta \gg \xi$.

\begin{figure}[h!]
  \includegraphics[width=85mm]{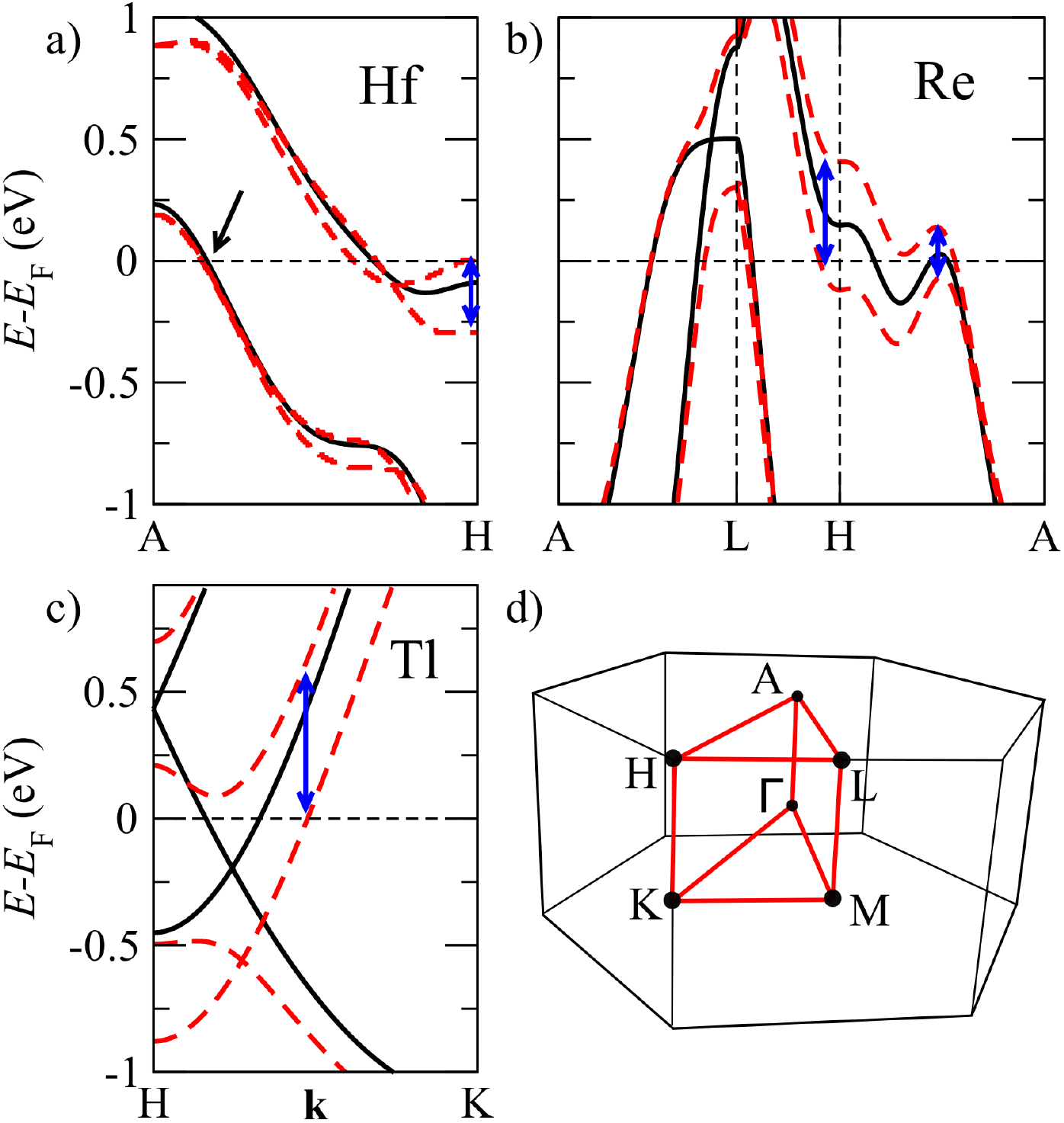}%
  \caption{(color online) a)-c): Band structures without (black solid lines) and with (red dashed lines) SOC along high symmetry lines in the Brillouin zone. Blue double-arrows denote a large SOC splitting of bands. A black arrow indicates a band in Hf which exhibits only a small SOC splitting. In panel d) the hcp Brillouin zone with high-symmetry points is shown.}
  \label{fig:bandstructures}
\end{figure}

For hcp-crystals, the uniaxiality is evident. Now, we exemplify the importance of a four-fold quasi degeneracy with a detailed study of the selected hcp-crystals Hf, Re and Tl by band-structure calculations. Without SOC, the spin-degeneracy in a non-magnetic crystal for every state is obvious. Additionally, an orbital degeneracy is always present for states where $\vcc{k}$ is on the hexagonal face of the Brillouin zone boundary (\cf black solid lines in Fig.~\ref{fig:bandstructures}). This special feature of the hcp crystal-structure is enforced by symmetry of their non-symmorphic space group \cite{Ashcroft_Mermin}. Similarly, the states on the high-symmetry line $H$-$K$ in Tl and Hf are four-fold degenerate (as we see in Figs.~\ref{fig:bandstructures}c and \ref{Fig:bandsHfBfield}a). These degeneracies are lifted due to SOC into two pairs (red dashed lines in Fig.~\ref{fig:bandstructures}), and hence the necessary conditions to form a spin-flip hot loop at the hexagonal face of the BZ are always fulfilled. Since the bands extend above and below the Fermi level on a large energy scale of 1~eV or more, the effect that we describe will be stable with respect to pressure, doping, or temperature. The SOC-splitting depends on the band index and $\vcc{k}$-point in the BZ, and can be as large as $0.5~\mathrm{eV}$ (\eg Re or Tl, see blue double-arrows in Fig.~\ref{fig:bandstructures}b-c). For these bands, the condition $\delta \lesssim \xi$ is not only fulfilled if the states are degenerate, but also if they are split up to some $\delta \lesssim 0.5~\mathrm{eV}$ in the absence of SOC.

Such a splitting $\delta$ occurs if $\vcc{k}$ departs from the hexagonal face towards the interior of the BZ and will grow with distance. Eventually, it will exceed the value $\xi$, and at this point $\bsqks$ decays and the the spin-flip hot loop ends.
With these arguments, we can explain the different thickness of spin-flip hot loops near the hexagonal face of the BZ, best seen in Hf: the band with the large SOC-splitting near the $H$-point (see blue double-arrow in Fig.~\ref{fig:bandstructures}a) forms the outer Fermi-surface sheet, which develops thick spin-flip hot loops for an SQA in the $ab$-plane (\cf Fig.~\ref{fig:FSgallery_hcp}). On the other hand, the band that crosses the Fermi level closer to the $A$-point (see black single arrow in Fig.~\ref{fig:bandstructures}a) is much weaker SOC-split and develops only a 100 times thinner loop on the inner FS-sheet (hardly visible in Fig.~\ref{fig:FSgallery_hcp}).

\subsection{Influence of an external $B$-field}

An additional orbital degeneracy to the conjugation degeneracy in absence of SOC is a prerequisite for the occurrence of spin-flip hot spots. In the previous paragraph we analyzed how quickly $\bsqks$ decays when the initial fourfold degeneracy on a high-symmetry line is broken by moving to $k$-points away from this line. Another way to lift the initial fourfold degeneracy is through breaking of conjugation symmetry by an external $B$-field of the form $\vcc{B} \cdot \pauli$ \footnote{We remind that conjugation symmetry is the combined action of space-inversion and time-reversal symmetry. The latter is broken by the external $B$-field.}.

The spin-quantization axis (SQA) is necessarily aligned parallel to the $B$-field. By rotating the SQA, we change the spin-conserving part, $\xi (LS_\parallel)$, and spin-flip parts of SOC, $\xi (LS^{\uparrow \downarrow})$ \cite{Mokrousov:Review:2013}. As an example, for $\sqa \parallel z$ the spin-conserving part reads $\xi L_z \, S_z$, whereas for $\sqa \parallel x$ it changes to $\xi L_x \, S_x$. Evidently, the spin-conserving part of SOC couples bands of the \emph{same} spin-character, whereas the spin-flip part couples those of \emph{opposite} spin-character.

\begin{figure}[b!]
  \centering
    \includegraphics[width=85mm]{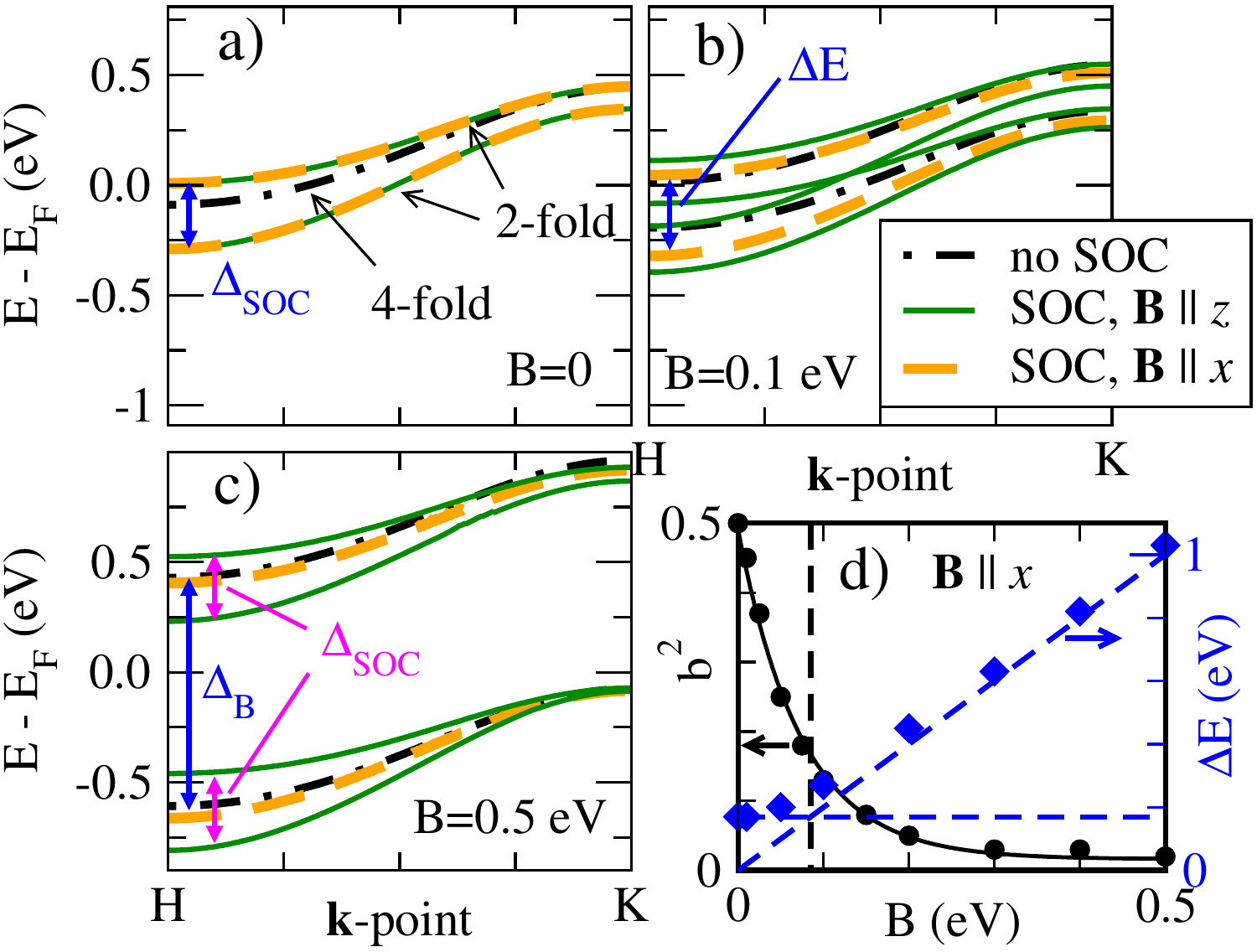}
  \caption[SOC-effects on the band structure of Hf under an applied $B$-field.]{(color online) a)-c): The splitting of the 4-fold degeneracy in presence of SOC and an exchange $B$-field is shown for Hf along the path $H$-$K$ for $B=0,\,0.1~\mathrm{and}~0.5\,\mathrm{eV}$. Dash-dotted lines represent bands without SOC, full and dashed lines with SOC and $\sqa$ along $z$ and $x$, respectively. In d), the spin-mixing parameter $b^2$ (circles) and splitting of energy bands (diamonds) for a selected $k$-point on the path $H$-$K$ and $\vcc{B} \parallel \sqa=x$ as a function of the field strength $B$ is shown. A strong decrease of $b^2$ with increasing $B$ is observed. The black solid line is an exponential fit (see text for details). The type of energy splitting is changed from SOC-dominated (denoted by the horizontal blue-dashed line at $\Delta_\mathrm{SOC} = \, 170\mathrm{meV}$) to $B$-dominated (denoted by the diagonal blue-dashed line $\Delta E = 2B$). The crossover is defined as the point where $2B = \Delta_\mathrm{SOC}$, and is indicated by the vertical black-dashed line.}
  \label{Fig:bandsHfBfield}
\end{figure}

As we explain in the following, through the increase of the strength of $B$ we are able to reduce the effect of the spin-flip part when SOC is added to the non-relativistic band structure, whereas the spin-conserving part remains at its full strength. We analyze for various $B$ the change of the band structure upon the inclusion of SOC in the high-symmetry path $H$-$K$ for hcp Hf (\cf Fig.~\ref{Fig:bandsHfBfield}).

Let us first consider a vanishing $B$-field (\cf Fig.~\ref{Fig:bandsHfBfield}a): the non-relativistic bands are fourfold degenerate due to the conjugation and the orbital symmetry. Then, the non-symmorphic degeneracy is lifted by SOC into two twofold degenerate pairs. On the one hand, the splitting $\Delta_\mathrm{SOC}$ is caused by the spin-conserving part if the SQA is along the $z$-axis, and on the other hand the same splitting is caused by the spin-flip part for a SQA along $x$. This fact was checked by separately acting with the spin-conserving and spin-flip part of $\vcc{L} \cdot \vcc{S}$, respectively, when calculating the band structure of Fig.~\ref{Fig:bandsHfBfield}a (not shown).

We now apply a strong $B$-field of $0.5\,\mathrm{eV}$ (\cf Fig.~\ref{Fig:bandsHfBfield}c), first without SOC: the bands are split into a pair of spin-up and a pair of spin-down states, with an energy difference between the pairs of $\Delta_B = 2B$. It is important to note that the states with the \emph{same} spin-character remain degenerate, and the situation is independent on the direction of $\vcc{B}$ as SOC was neglected.
We now include SOC, first for a $\vcc{B}$-field (and thus the SQA) along $z$: we observe that each degenerate pair acquires a full SOC-splitting $\Delta_\mathrm{SOC}$, of same size as for the case $B=0$ (compare splittings solid lines in Figs.~\ref{Fig:bandsHfBfield}a and c). The conclusion is that this splitting must be fully governed by the spin-conserving part of SOC (the same conclusions can be drawn for a smaller $\vcc{B}$-field of 0.1~eV, see Fig.~\ref{Fig:bandsHfBfield}b).
Next, we analyze the response of the non-relativistic degenerate pairs upon inclusion of SOC for $\vcc{B}$ (and SQA) along $x$. Nearly no response of the bands is observed (see yellow dashed lines in Fig.~\ref{Fig:bandsHfBfield}c).  

Clearly, now the spin-conserving part of SOC is ``deactivated'' for these particular bands, in strong contrast to the case that $\vcc{B} \parallel z$. The question is: was the spin-flip part at the same time activated? At this large $B$ field, we are not able to judge it, as  it couples states of different spin-character, which are separated by a rather large energy of $2B=1\,\mathrm{eV}$, and thus the effect of spin-flip SOC is strongly suppressed. For intermediate $\vcc{B}$-fields (\cf dashed line in Fig.~\ref{Fig:bandsHfBfield}b), the degeneracy is still present because both states in the up-band couple to the states in the down-band the same way due to their special symmetry. In the limit $B=0$, the full SOC-splitting is restored, but now caused by the spin-flip part (see above).

A detailed analysis of the band splitting as a function of $B$ (for $\vcc{B} \parallel x$) is shown in Fig.~\ref{Fig:bandsHfBfield}d for a selected $\vcc{k}$-point in $H$-$K$. With decreasing $B$, a crossover between $B$-field dominated and SOC-dominated regimes can be seen, which is well characterized by the condition $2B = \Delta_\mathrm{SOC}$ (indicated by a vertical line in Fig.~\ref{Fig:bandsHfBfield}d). Simultaneously, the spin-mixing parameter of the states, $b^2$, decreases exponentially from its maximal value of $0.5$ at vanishing $B$ towards a value $b^2_\infty$ for large $B$, described by the function $b^2(B) = (\frac{1}{2}-b^2_\infty)\, \expo{-2B/\Delta}+b^2_\infty$. We fitted parameters $b^2_\infty = 1.95 \times 10^{-2}$ and $\Delta=145~\mathrm{meV}$ (see black solid line in Fig.~\ref{Fig:bandsHfBfield}d), which is in good agreement to the spin-orbit splitting $\Delta_\mathrm{SOC}=170~\mathrm{meV}$.

This analysis allows us as well to estimate the stability of the spin-flip hot loops against an external magnetic field: to significantly decrease the intensity of the loop by a factor of $1/2$, a field about 50~meV is necessary (which corresponds to about 1~kTesla). This is similar to findings in magnetic materials \cite{Haag_Faehnle:2014}, where the magnitude of the EYP is stable up to fields of the order of a few kTesla. Such large fields could probably be produced by a proximity effect to a ferromagnet, but only in the first few interface layers, so it is unlikely that they occur in bulk.

Similarly, we expect that the predicted anisotropy should remain finite even at room temperature, where a broadening of the bands on the order of 25meV is smaller than the spin-orbit splitting. The situation could be different for both perturbations, external $B$-field and temperature effects, if SOC is much weaker.

\subsection{$3d$ and $4d$ metals with hcp crystal structure}

We turn our attention to hcp crystals with smaller SOC, namely the 4$d$ elements Y, Zr, Tc, Ru and Cd, and the 3$d$ (non-magnetic) crystals of Sc, Ti and Zn. Additionally we consider the very light (small SOC) element Mg.

Our results are collected in Table~\ref{table:EYP_all}. 
They show that the anisotropy of $b^2$ can reach colossal values of up to $6000\%$ for Ti compared to $830\%$ for Hf, or $1250\%$ for Sc compared to 200\% for Lu, which are iso-electronic to each other. Generally speaking, we observe the trend that the anisotropy increases from $5d/6sp$ elements to $4d$ elements to $3sp/3d$ elements. This comes as a surprise, since from a decrease in SOC strength also a decrease of the anisotropy could have been expected. However, as our calculations show and we analyze further in the rest of this section, the anisotropy increases for light elements due to the presence of spin-flip hot loops. In the following, we discuss some of these metals in more detail.

\begin{figure*}[htbp!]
  \includegraphics[width=170mm]{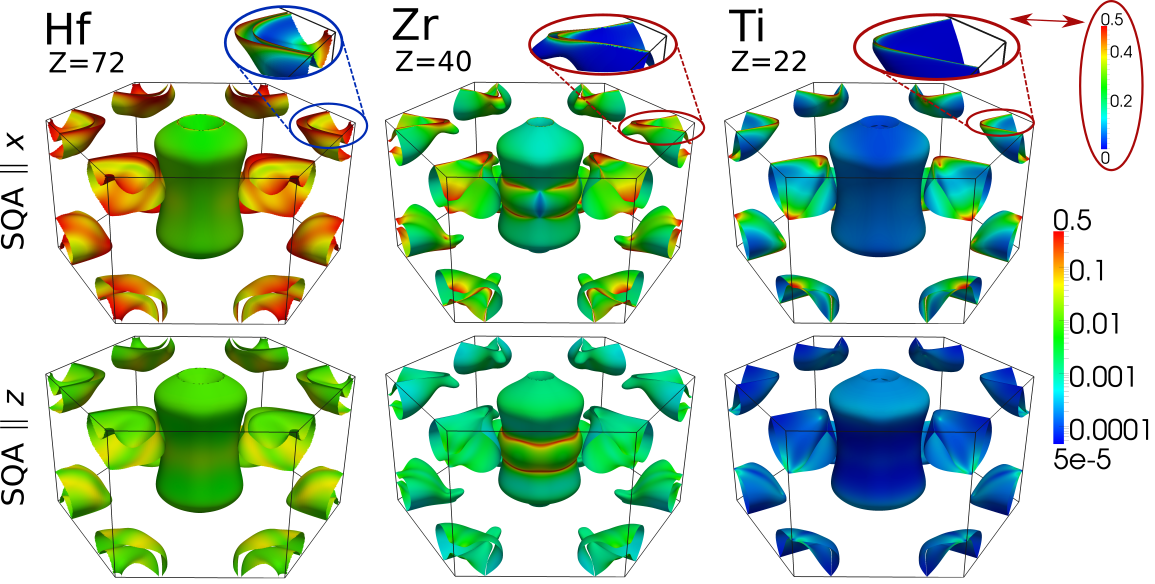}%
  \caption{(color online) Spin-mixing parameter $\bsqks$ for the iso-electronic elements Hf, Zr and Ti ($Z$ is the atomic number) for two directions of the SQA. The weakening of the spin-orbit coupling strength when going from Hf via Zr to Ti results in a smaller $\bsqks$ in most parts of the Fermi surface. Also the width of the spin-flip hot loops becomes smaller, but they remain finite even for Ti. The insets display the loop region with a linear color scale. \label{Fig:FSgalleryTiZrHf}}
\end{figure*}

We investigate this trend in detail by examining the elements Hf, Zr and Ti. This trio of elements is well suited for a study of the influence of the spin-orbit coupling strength because (i) they all crystallize in the hcp crystal structure, (ii) their iso-electronic valence band configuration results in very similar Fermi surfaces, and (iii) the SOC strength $\xi$ varies by about one order of magnitude ($\xi_\mathrm{Hf}/\xi_\mathrm{Ti} \approx Z_\mathrm{Hf}^2 / Z_\mathrm{Ti}^2 \approx 10$, where $Z$ is the atomic number of the respective element; see detailed analysis below).

\begin{figure}
  \includegraphics[width=85mm]{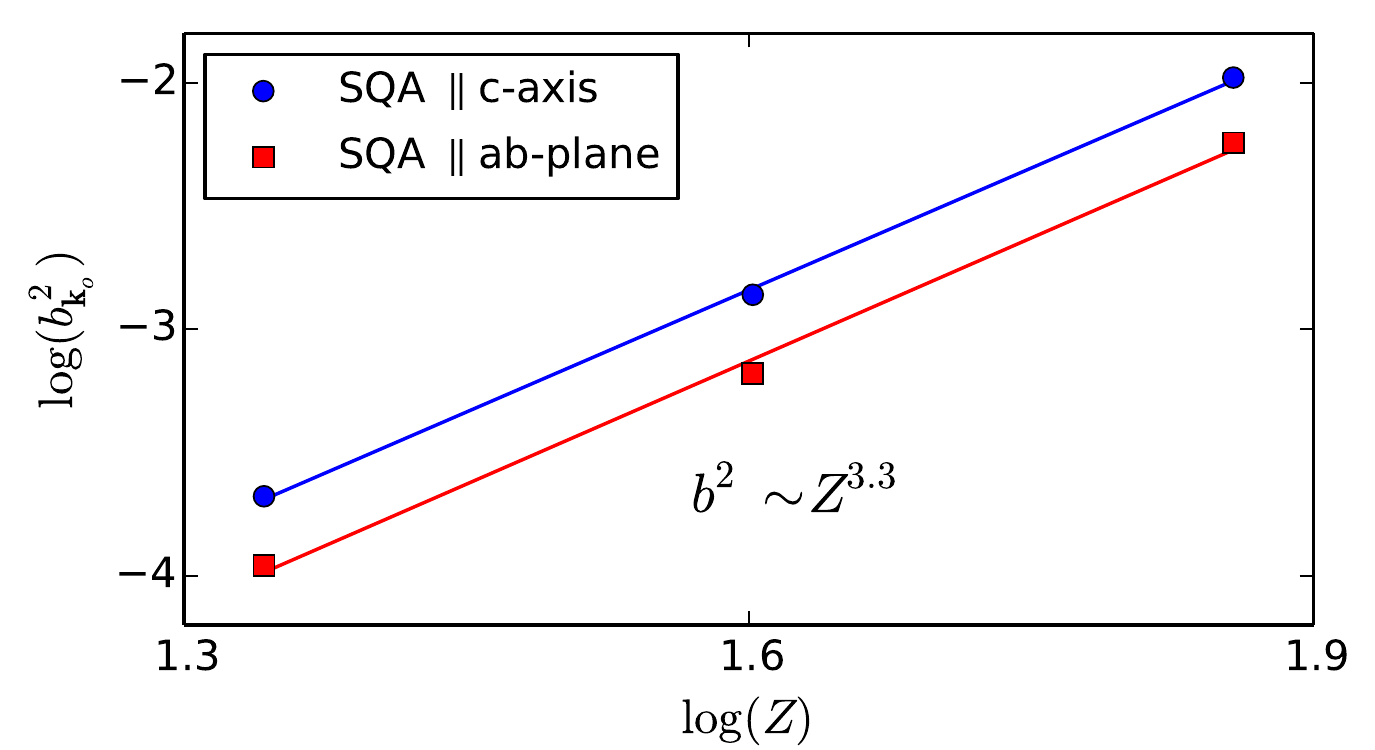}%
  \caption{(color online) Scaling of $\bsqks$ at a selected $k$-point on the inner sheet of the Fermi surfaces of Ti, Zr and Hf as a function of the atomic number. The least squares fit yields $\bsqks \sim Z^{3.3}$, independent on the direction of the SQA. \label{Fig:bsqfit}}
\end{figure}

Let us compare the spin-mixing parameter on the Fermi surfaces of these three elements (see Fig.~\ref{Fig:FSgalleryTiZrHf}) with special attention to its scaling as a function of the atomic number $Z$. It is insightful to divide the Fermi surface into two parts: (i) ``ordinary'' regions, that do not exhibit a spin-flip hot spot and (ii) regions with spin-flip hot spots.
\begin{enumerate}[label=(\roman*)]
	\item{At an ordinary point on the Fermi surface, \eg the central Fermi surface sheet in Fig.~\ref{Fig:FSgalleryTiZrHf}, $b_{\vcc{k}}^2$ decreases from Hf via Zr to Ti as inferred from the color on the logarithmic scale. A detailed analysis for a selected, ordinary $\vcc{k}$-point reveals that the spin-mixing parameter scales as $b^2_{\vcc{k}} \sim Z^{3.3}$ (see Fig.~\ref{Fig:bsqfit})\footnote{This allows us to approximate for the scaling of the spin-orbit coupling strength $\xi \sim Z^{1.65}$ (since $b^2 \sim \xi^2$ according to Elliott \cite{Elliott.1954}).}. Hence, the contribution  of these ordinary regions to the Fermi-surface average in Ti is smaller by two orders of magnitude compared to Hf.}
	\item{In contrast, the spin-flip hot-loop at the hexagonal face of the BZ in Hf for $\sqa \parallel ab$~plane remains present also for Zr and Ti (see upper part of Fig.~\ref{Fig:FSgalleryTiZrHf}). The thickness of the hot loop decreases from Hf via Zr to Ti because of the smaller SOC strength, but importantly $b^2_{\vcc{k}}=0.5$ remains at its maximal value directly on the hexagonal face. As a result, the Fermi-surface average is dominated by this contribution and scales very different with $Z$ (roughly as $Z^2$). Hence it is only one order of magnitude smaller in Ti compared to Hf.}
\end{enumerate}

\begin{figure*}[htbp!]
	\includegraphics[width=0.8\textwidth]{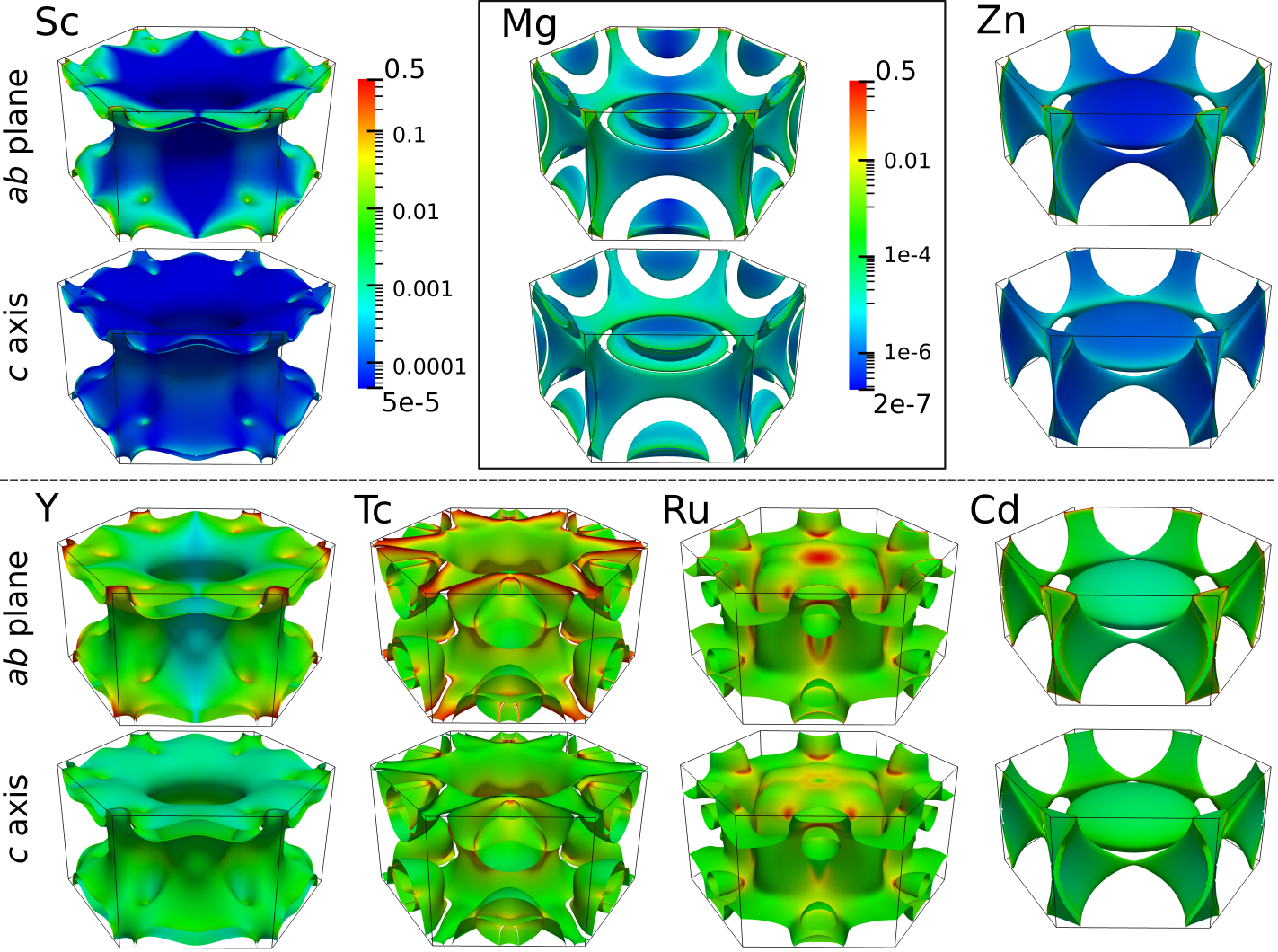}%
	\caption{(color online) Spin-mixing parameter $\bsqks$ as color code on the Fermi surface for various hcp-elements. The top and bottom pictures of each element correspond to two different directions of $\sqa$. For all the plots (except Mg) the same color legend is used.\label{Fig:lightHCPgallery}}
\end{figure*}

To summarize, the fact that the spin-flip hot loops remain at their maximal value directly on the hexagonal face of the BZ, that they disappear for one direction of $\sqa$, and that the remainder of the FS gives almost no contribution causes the colossal anisotropy of $b^2$ in Ti. This trend is also well observed for the three iso-electronic elements Sc, Y and Lu (see Figs.~\ref{fig:FSgallery_cub} and \ref{Fig:lightHCPgallery}), and a steady increase of $\mathcal{A}$ with decreasing atomic number $Z$ is obtained (see Table~\ref{table:EYP_all}).

The drastic increase of the averaged EYP due to the presence of very thin spin-flip hot loops is qualitatively similar to the increase of the EYP in fcc-Al due to the presence of very small spin-flip hot spots, as found by Fabian and Das Sarma \cite{Fabian:HotSpots:1998}. In their study, the realistic calculation was compared to a fictitious one, where Al was modeled as a monovalent metal leading to a disappearance of spin hot-spots. In contrast, in our study of Ti (and many other hcp-crystals) we can make the spin-flip hot loops disappear by merely changing the SQA due to the particular conditions met in these hcp crystals.

The trend of increasing $\mathcal{A}$ with decreasing $Z$ is violated between Hf and Zr. This peculiarity can be attributed to two anomalies in the band structure: First, the anisotropy in Hf is \emph{enhanced} due to the fact, that in Hf a Fermi-surface sheet occurs to be close to the corner $H$-point of the BZ, which causes the Fermi velocity of this band to vanish and enhances the weight of this particular band in the integral in Eq.~\eqref{integral}. Since this precise band incorporates a broad spin-flip hot loop, the anisotropy is also enhanced from about $500\%$ to the reported value of $830\%$ \cite{Zimmermann:Spinrelaxation:2012}. Secondly, the anisotropy in Zr is \emph{reduced} due to the presence of a spin-flip hot-loop on the central sheet of the Fermi surface due to an accidental degeneracy of bands. This inner loop appears irrespective of the direction of the SQA and enhances the value of $b^2_{\sqa \parallel z}$ by a factor of $3-4$ compared to the ordinary scaling (according to Fig.~\ref{Fig:bsqfit}), which leads to a reduction of $\mathcal{A}$ in Zr.

\section{Conclusions\label{Sec:Conclusions}}

We have developed a tetrahedron based algorithm within the relativistic Korringa-Kohn-Rostoker Green function method for the accurate calculation of Fermi surfaces of very complicated shape, as frequently found for transition-metal crystals. We applied it to all 5$d$ metals (La, Lu, Hf, Ta, W, Re, Os, Ir and Pt), some 6$sp$ metals (Au, Tl and Pb), and selected lighter elements with hcp crystal structure (Mg, Sc, Ti, Zn, Y, Zr, Tc, Ru and Cd). Even fine features, such as small splittings of Fermi-surface sheets, which are frequently found in crystals of light elements due to their small spin-orbit coupling, are properly described.

We investigated the spin-mixing parameter, which is related to spin-relaxation of conduction electrons via the Elliott-Yafet mechanism, and in particular concentrated on its recently discovered anisotropy with respect to the spin-polarization direction of electrons \cite{Zimmermann:Spinrelaxation:2012}. Our scan through the $5d$ and $6sp$ metals shows, that hcp crystals exhibit in general a giant anisotropy of about 100\%, as opposed to cubic crystals with anisotropies of up to merely 1\%. Exceptions with an above-average anisotropy are hcp-Hf (830\%) and bcc-W (6\%). We identified the emergence of spin-flip hot loops at the hexagonal face of the hcp Brillouin zone as the main source for a giant anisotropy. We found that these hot loops (and consequently the anisotropy) should be stable under an external $B$-field with strength equivalent to the SOC splitting (which corresponds to about 1-2 kTesla in these $5d$ metals), as well as under moderate variations of the Fermi energy through pressure, temperature, or doping. We showed that through a large variation of $B$, the spin-mixing parameter $b^2$ at a spin-flip hot-spot can be tuned.

For light elements with hcp crystal structure, we find even higher anisotropies as compared to $5d$ hcp-crystals, reaching a colossal value for hcp-Ti of $6000\%$. Again, spin-flip hot-loops were identified as the main source. Due to the smaller SOC, they are thin (but of finite width), which leads in combination with a tiny spin-mixing parameter in the other parts of the BZ (scaling as $b^2 \propto Z^{3.3}$) to a colossal anisotropy.

Our calculations identify the light hcp crystals of Mg, Sc, Ti and Zn as promising materials for new spintronics applications, because the low Elliott-Yafet parameter of the order of $10^{-4}$ might enable long enough spin-diffusion lengths for real devices, and at the same time exhibits the largest anisotropies that can be exploited to tailor the spin-diffusion length. As an outlook, the inclusion of the explicit scattering mechanism via \eg impurities or phonons is necessary to make quantitative predictions of the spin-diffusion length and stability against temperature effects.

\begin{acknowledgments}
 We are indebted to J.~Fabian for an introduction to the field and for discussions. We are also indebted to R.~Zeller and P.~H.~Dederichs for their invaluable help in the KKR formalism and to S.~Heers, D.~S.~G.~Bauer, M.~Gradhand and P.~Baumeister for discussions. We acknowledge funding under DFG project No.\ SPP-1538 ``Spin Caloric Transport'', and HGF-YIG program VH-NG-513 as well as computing time at the J\"{u}lich Supercomputing Centre and JARA-HPC of RWTH Aachen University.
\end{acknowledgments}

\appendix

\section*{Appendix: Linear combination of conjugation-degenerate states: discussion and physical interpretation}
\label{appendix:linear_combination}

Assuming that the crystal Hamiltonian is invariant under the action of 
the operators of space-inversion (parity) $P: \boldsymbol{\Psi}(\vcc{r})\rightarrow \boldsymbol{\Psi}(-\vcc{r})$ and time-reversal $K: \boldsymbol{\Psi}(\vcc{r})\rightarrow i\pauliy \boldsymbol{\Psi}^*(\vcc{r})$, then for every Bloch eigenstate $\boldsymbol{\Psi}_{\vcc{k}}$ there exists a conjugate partner eigenstate $PK\boldsymbol{\Psi}_{\vcc{k}}$ that is degenerate at the same $\vcc{k}$, orthogonal to $\boldsymbol{\Psi}_{\vcc{k}}$, and has the opposite spin expectation value \cite{Yafet.1963}. This is the case for the pair of Eqs.~(\ref{Eq:eyaf:wavefun1},\ref{Eq:eyaf:wavefun2}). Practically, time reversal symmetry means absence of external or internal magnetic fields in the crystal Hamiltonian.

In the case of conjugation degeneracy, and given a Bloch eigenfunction $\boldsymbol{\Psi}_{\vcc{k}}$ of the crystal Hamiltonian, any linear combination $\boldsymbol{\Psi}_{\vcc{k}}$ and $PK\boldsymbol{\Psi}_{\vcc{k}}$ is obviously again a Bloch eigenfunction at the same $\vcc{k}$. For the particular problem of spin relaxation, correspondence to experiment leads us to choose a linear combination that maximizes the spin expectation value along \sqa. For this we use the unitary transformation
\begin{eqnarray}
\boldsymbol{\Psi}_{\vcc{k} \sqa}^{+} &=& \cos\theta_{\vcc{k}}  \boldsymbol{\Psi}_{\vcc{k}} + \sin\theta_{\vcc{k}} e^{-i\phi_{\vcc{k}}} PK \boldsymbol{\Psi}_{\vcc{k}} \\
\label{Eq:eyaf:wavefun3}
\boldsymbol{\Psi}_{\vcc{k} \sqa}^{-} &=& -\sin\theta_{\vcc{k}} e^{i\phi_{\vcc{k}}} \boldsymbol{\Psi}_{\vcc{k}} + \cos\theta_{\vcc{k}} PK \boldsymbol{\Psi}_{\vcc{k}} 
\label{Eq:eyaf:wavefun4}
\end{eqnarray}
where two real parameters on the Bloch sphere, $\theta_{\vcc{k}}\in[0,\pi]$ and $\phi_{\vcc{k}}\in[0,2\pi]$, suffice for the definition of the normalized linear combinations up to an arbitrary global phase.
It is easy to see that $\boldsymbol{\Psi}_{\vcc{k} \sqa}^{-}=PK\boldsymbol{\Psi}_{\vcc{k} \sqa}^{+}$. Following Refs.~\cite{Elliott.1954,Fabian:HotSpots:1998} we then define appropriate $\theta_{\vcc{k}}$ and $\phi_{\vcc{k}}$ such that $S^{+}_{\vcc{k}\sqa}:= \bra{\boldsymbol{\Psi}_{\vcc{k} \sqa}^{+}} \frac{\hbar}{2}\pauli\cdot\sqa\ket{\boldsymbol{\Psi}_{\vcc{k} \sqa}^{+}}\geq 0$ is maximal or equivalently  $S^{-}_{\vcc{k}\sqa}:= \bra{\boldsymbol{\Psi}_{\vcc{k} \sqa}^{-}} \frac{\hbar}{2}\pauli\cdot\sqa\ket{\boldsymbol{\Psi}_{\vcc{k} \sqa}^{-}}=-S^{+}_{\vcc{k}\sqa}\leq 0$ is minimal (maximal in absolute value). By demanding the derivatives with respect to $\theta$ and $\phi$ to vanish, we obtain a maximal $S^{+}_{\vcc{k}\sqa}$ under the condition:
\begin{equation}
 \phi_{\vcc{k}} = \arg \left( S_{12}^{\sqa} \right), ~~ \theta_{\vcc{k}} =\frac{1}{2} \atan \,  \frac{|S_{12}^{\sqa}|}{ S_{1}^{\sqa} }, \label{eq:app:Szmax}
\end{equation}
where $S_1^{\sqa}=\frac{\hbar}{2}\bra{\boldsymbol{\Psi}_{\vcc{k}}}\pauli\cdot\sqa\ket{\boldsymbol{\Psi}_{\vcc{k}}}$ and $S_{12}^{\sqa}=\frac{\hbar}{2}\bra{\boldsymbol{\Psi}_{\vcc{k}}}\pauli\cdot\sqa\ket{PK\boldsymbol{\Psi}_{\vcc{k}}}$. We should point out that this condition is equivalent to imposing $\bra{\boldsymbol{\Psi}_{\vcc{k} \sqa}^{-}}\pauli\cdot\sqa\ket{\boldsymbol{\Psi}_{\vcc{k} \sqa}^{+}}=0$, as shown in Ref.~\cite{Pientka.2012}.

This particular selection of $\theta_{\vcc{k}}$ and $\phi_{\vcc{k}}$, \ie\ maximizing the spin expectation value along \sqa, is of course just one of (infinitely) many possibilities, each of them corresponding, in principle, to the idealization of a different experimental setup. The particular one is motivated by the fact that in conduction electron spin resonance experiments, an external magnetic field $\vcc{B}_{\rm ext}$ lifts the conjugation degeneracy, modeled by the Hamiltonian $\vcc{B}_{\rm ext}\cdot \pauli$. Since the field is weak compared to the interband separation we may solve the eigenvalue problem in the subspace spanned by $\boldsymbol{\Psi}_{\vcc{k}}$ and $PK\boldsymbol{\Psi}_{\vcc{k}}$ arriving at the aforementioned condition. Our results in the present paper are calculated using this condition \eqref{eq:app:Szmax}.

A second condition that has been chosen in the past \cite{Gradhand:Spinmixing:2010} is to demand that $\bra{\boldsymbol{\Psi}_{\vcc{k} \sqa}^{+}} \frac{\hbar}{2}\pauli\cdot\sqa'\ket{\boldsymbol{\Psi}_{\vcc{k} \sqa}^{+}}=\bra{\boldsymbol{\Psi}_{\vcc{k} \sqa}^{+}} \frac{\hbar}{2}\pauli\cdot\sqa''\ket{\boldsymbol{\Psi}_{\vcc{k} \sqa}^{+}}=0$, where $\sqa'$ and $\sqa''$ form together with \sqa\ an orthonormal reference system (\eg the $x,y,z$ axes). This choice implies that the states $\boldsymbol{\Psi}_{\vcc{k} \sqa}^{\pm}$ have a spin projection purely along the \sqa\ axis (which was not the case for the first condition), and could plausibly represent an experiment where electrons with selected spin strictly along \sqa\ are injected into a material from the outside, so that they have to be accommodated by Bloch states also without perpendicular spin components. One then obtains different equations for $\theta_{\vcc{k}}$ and $\phi_{\vcc{k}}$ \cite{Heers:82648,Gradhand:Spinmixing:2010}.

The previous two conditions give very similar values for $S^{+}_{\vcc{k}\sqa}$, except in the case that $S^{+}_{\vcc{k}\sqa}$ becomes small, \ie\ close to spin-flip hot spots; for this reason, the anisotropy of the spin relaxation is also different, although it is of the same order of magnitude. 

Pientka \etal \cite{Pientka.2012} call this a choice of gauge.
Concerning spin relaxation, they find that the two choices give similar but slightly different results for the case of impurity scattering in Cu, Ag, Pt. The same conclusion on spin relaxation was reached by Heers \cite{Heers:82648}.

A third condition has been introduced by Long \etal \cite{Long:W001:2013} for the calculation of the surface Rashba states in thin films. Here the degeneracy to be lifted is related to choosing between two degenerate surface states located at two opposite film surfaces; the related experiment would be an electron injection or a scanning tunneling measurement at one surface. The choice of $\theta$ and $\phi$ is such that the charge or the spin expectation value is maximized on one surface and in the vacuum region adjacent to it.

Generally speaking, any condition lifting the degeneracy in the $\{\boldsymbol{\Psi}_{\vcc{k}},PK \boldsymbol{\Psi}_{\vcc{k}}\}$ subspace specifies a basis $\{ \boldsymbol{\Phi}_{\vcc{k}}=U\boldsymbol{\Psi}_{\vcc{k}},\bar{\boldsymbol{\Phi}}_{\vcc{k}}=PK \boldsymbol{\Phi}_{\vcc{k}}\}$, where $U$ is a unitary transformation in the $2\times2$ subspace, and thereby represents a specific observable (defined mathematically by its eigenvectors $\boldsymbol{\Phi}_{\vcc{k}}$, $\bar{\boldsymbol{\Phi}}_{\vcc{k}}$) and corresponds to a unique type of measurement. To this point, some insight can be gained by the following observation. Considering the
spin polarization $\vcc{S}_{\boldsymbol{\Phi}_{\vcc{k}}}:=\frac{\hbar}{2}\bra{\boldsymbol{\Phi}_{\vcc{k}}}\pauli\ket{\boldsymbol{\Phi}_{\vcc{k}}}$ and the corresponding unit vector $\sqa_{\boldsymbol{\Phi}_{\vcc{k}}}:=\vcc{S}_{\boldsymbol{\Phi}_{\vcc{k}}}/|\vcc{S}_{\boldsymbol{\Phi}_{\vcc{k}}}|$, it is obvious that the projection $\vcc{S}_{\boldsymbol{\Phi}_{\vcc{k}}}\cdot \sqa_{\boldsymbol{\Phi}_{\vcc{k}}}$ is maximal with respect to all possible projections of the type $\vcc{S}_{U\boldsymbol{\Phi}_{\vcc{k}}}\cdot \sqa_{\boldsymbol{\Phi}_{\vcc{k}}}$, since $U$ will mix in terms containing $\bar{\boldsymbol{\Phi}}_{\vcc{k}}$ that is characterized by the opposite spin (see \eg Ref.~\cite{Pientka.2012} for a proof). But this means that the pair $\boldsymbol{\Phi}_{\vcc{k}}$ and $\bar{\boldsymbol{\Phi}}_{\vcc{k}}$ is just the pair $\boldsymbol{\Psi} _{\vcc{k}}^{\pm}$ that maximizes the spin in the direction $\sqa_{\boldsymbol{\Phi}_{\vcc{k}}}$, as defined by the first condition previously; i.e., the pair $\boldsymbol{\Phi}_{\vcc{k}}$ and $\bar{\boldsymbol{\Phi}}_{\vcc{k}}$ defines the eigenstates of a perturbation by a Zeeman magnetic field along $\sqa_{\boldsymbol{\Phi}_{\vcc{k}}}$. Therefore, any condition lifting the degeneracy can be physically seen as imposing a $\vcc{k}$-dependent magnetic field defined in this way. The first condition, used throughout the present paper, merely corresponds to the special case of a $\vcc{k}$-independent field.

An application of the concept of a $\vcc{k}$-dependent Zeeman field is the celebrated spin-orbit field observed in systems with lifted space-inversion symmetry \eg in the conduction band of zinc-blende or wurtzite structure semiconductors \cite{PhysRev.100.580} or in the band structure of noble-metal surface states \cite{Henk_PRB_Au111_spinpol,LaShell_Au111_spinsplitting}. Here, the observable that breaks the symmetry is the anti-symmetric part $V_{\rm A}$ of the crystal potential $V$ that can be written with the help of the parity operator as $V_{\rm A}=\frac{1}{2} (V-PVP^{-1})$. Let $\boldsymbol{\Psi}_{\vcc{k}}$ and $\bar{\boldsymbol{\Psi}}_{\vcc{k}}=PK \boldsymbol{\Psi}_{\vcc{k}}$ be degenerate conjugate Bloch eigenstates corresponding to the symmetric part $V_{\rm S}=V-V_{\rm A}$, but otherwise arbitrarily chosen within the $2\times2$ conjugate subspace. Then $V_{\rm A}$ causes a lifting of degeneracy through the Hamiltonian
\begin{equation}
H_{\vcc{k}}=\left(
\begin{array}{cc}
\bra{\boldsymbol{\Psi}_{\vcc{k}}}V_{\rm A}\ket{\boldsymbol{\Psi}_{\vcc{k}}} & \bra{\boldsymbol{\Psi}_{\vcc{k}}}V_{\rm A}\ket{\bar{\boldsymbol{\Psi}}_{\vcc{k}}} \\
\bra{\bar{\boldsymbol{\Psi}}_{\vcc{k}}}V_{\rm A}\ket{\boldsymbol{\Psi}_{\vcc{k}}} &  \bra{\bar{\boldsymbol{\Psi}}_{\vcc{k}}}V_{\rm A}\ket{\bar{\boldsymbol{\Psi}}_{\vcc{k}}}
\end{array}
\right).
\end{equation}
Since the asymmetric potential satisfies $V_{\rm A}P=-PV_{\rm A}$, it is straightforward to show that $\bra{\bar{\boldsymbol{\Psi}}_{\vcc{k}}}V_{\rm A}\ket{\bar{\boldsymbol{\Psi}}_{\vcc{k}}}=-\bra{\boldsymbol{\Psi}_{\vcc{k}}}V_{\rm A}\ket{\boldsymbol{\Psi}_{\vcc{k}}}$. Thus the Hamiltonian $H_{\vcc{k}}$ is traceless, producing a symmetric splitting $E_{\vcc{k}}^{\pm} = E_{\vcc{k}}\pm \hbar|\Omega_{\vcc{k}}|$, where $|\Omega_{\vcc{k}}| = \frac{1}{\hbar}\left[\,\bra{\boldsymbol{\Psi}_{\vcc{k}}}V_{\rm A}\ket{\boldsymbol{\Psi}_{\vcc{k}}}^2+|\bra{\boldsymbol{\Psi}_{\vcc{k}}}V_{\rm A}\ket{\bar{\boldsymbol{\Psi}}_{\vcc{k}}} |^2\,\right]^{1/2}$ is the magnitude of the spin-orbit field. The spin polarization of the two resulting eigenstates $\boldsymbol{\Psi}_{\vcc{k}}^{\pm}$, $\vcc{S}_{\vcc{k}}^{\pm}=\pm \frac{\hbar}{2}\bra{\boldsymbol{\Psi}_{\vcc{k}}^{+}}\pauli\ket{\boldsymbol{\Psi}_{\vcc{k}}^{+}}$, defines the direction of the spin-orbit field, yielding $\vcc{\Omega}_{\vcc{k}}:=|\Omega_{\vcc{k}}|\, \vcc{S}_{\vcc{k}}^+/|\vcc{S}_{\vcc{k}}^+|$. At the end, the vector $\hbar \vcc{\Omega}_{\vcc{k}}$ plays the role of the $\vcc{k}$-dependent Zeeman field, discussed in the previous paragraph, corresponding implicitly to the choice of degeneracy lifting through the asymmetry $V_{\rm A}$.

\bibliography{ref}

\end{document}